\documentclass[letterpaper]{JHEP3}
\usepackage[centertags]{amsmath}
\usepackage{amssymb}
\usepackage{graphicx}
\usepackage{color}

\def\l{\lambda}
\def\lb{\bar{\lambda}}
\def\e{\eta}
\def\>{\rangle}
\def\<{\langle}
\date{}
\title {The Noncommutative S-Matrix}
\author{Suvrat Raju \\ Harish-Chandra Research Institute, \\ Jhunsi, Allahabad 211019\\ E-mail: \email{suvrat@hri.res.in}}
\abstract{As a simple example of how recently developed on-shell techniques apply to nonlocal theories, we study the S-matrix of noncommutative gauge theories. In the complex plane, this S-matrix has essential singularities that signal the nonlocal behavior of the theory. 
In spite of this, we show that tree-level amplitudes may be obtained by BCFW type recursion relations. At one loop we find a complete basis of master integrals (this basis is larger than the corresponding basis in the ordinary theory). Any one-loop noncommutative amplitude may be written as a linear combination of these integrals with coefficients that we relate to products of tree amplitudes. We show that the noncommutative ${\mathcal N = 4}$ SYM theory has a structurally simple S-matrix, just like the ordinary ${\mathcal N = 4}$ SYM theory.}

\preprint{HRI/ST/0906}

\begin{document}
\section{Introduction}
The Feynman diagram expansion is a very inefficient approach to gauge theory amplitudes. We are often interested only in on-shell quantities but Feynman diagrams also carry a lot of off-shell information. 
On-shell scattering amplitudes in gauge theories can be remarkable simple. For example, the scattering of two gluons with negative helicity and any number of gluons
 with positive helicity 
is given by the compact Parke-Taylor formula \cite{Parke:1986gb}. This structure is entirely obscured by Feynman diagrams.

This difficulty is more than aesthetic. A modest increase in the number of external legs causes Feynman diagrams to proliferate.  This makes it difficult even for modern computers to calculate amplitudes for processes relevant at the LHC!

Starting in the early nineties Bern et al. \cite{Bern:1994cg,Bern:1994zx,Bern:1995db} 
developed efficient techniques to calculate one-loop amplitudes from on-shell tree level 
amplitudes using, what is called, `generalized unitarity'. Surprisingly on-shell techniques for tree level amplitudes came only later with the discovery of the BCFW recursion relations \cite{Britto:2004ap,Britto:2005fq}. However, it is now possible, starting with only the 
{\it on-shell, three point amplitude},  to calculate arbitrary one loop amplitudes in a gauge theory. This approach to amplitudes is not only efficient, it also lends itself to easy automation \cite{Berger:2008sj}. 

The current techniques work well for gauge theories at one loop but it is not unreasonable to hope that they are the nucleus of a reformulation of perturbative quantum field theory. Apart from computational simplicity, what can we hope to gain from this new perspective? One answer 
suggested in \cite{ArkaniHamed:2008gz} was that this approach might help us move 
away from locality!

Contrary to expectations based on the Lagrangian, \cite{ArkaniHamed:2008gz} pointed out that scattering amplitudes in ${\mathcal N}=4$ Super Yang-Mills (SYM) are `simpler'\footnote{The criterion of simplicity here is aesthetic rather than computational. See section \ref{problematizesimplicity}.} than amplitudes in pure Yang-Mills (YM) which, in turn, are simpler than amplitudes in a scalar theory. The authors of \cite{ArkaniHamed:2008gz} wondered if the Lagrangian formulation obscures this simplicity because it keeps locality manifest. This led them to suggest that one should look for a dual formulation of quantum field theory that would make this simplicity, rather than locality, explicit.

Even more ambitiously, one could ask whether this new approach helps us generalize the 
structure of local quantum field theory. While it is simple to write down a nonlocal
classical Lagrangian such an action almost always runs into trouble with unitarity. 
However, in the resuscitated S-matrix approach to quantum field theory, it is unitarity 
that is kept manifest rather than locality. This makes it a natural framework to study 
these questions.

Unfortunately, these arguments are deceptively simple. The 
S-matrix approach to quantum field theory relies heavily on the analytic properties of 
amplitudes. These in turn are tightly linked to locality (see, for instance, \cite{Weinberg:1995mt}). For example, consider a nonlocal
interaction smeared over a distance $l$. To represent this using a local Lagrangian,
we need an infinite sequence of derivatives --- very roughly, because $f(x+l) = e^{l {\partial \over \partial x}} f(x)$. In momentum space, this leads to an essential singularity in the complex plane. 
This is familiar from string theory. 
Recall that the Veneziano amplitude in the hard scattering limit goes like $\exp{\left(-\alpha' E^2\right)}$ (where $E$ is some energy scale describing the external particles) \cite{Polchinski:1998rq}.

So, it is of interest to study how well S-matrix techniques stand up when they are applied to field theories with novel analytic properties. This is what we do in this paper.

We will study scattering in noncommutative field theories. Noncommutative field theories are obtained by considering quantum field theory on a spacetime where the underlying coordinates do not commute
\begin{equation}
[x^{\mu}, x^{\nu}] = i \theta^{\mu \nu}.
\end{equation}
From the point of view of quantum field theory, this is equivalent to replacing the ordinary pointwise product of fields in the Lagrangian with the star product
\begin{equation}
 \left(f * g \right)(x) = \lim_{x \rightarrow y} e^{{i \over 2} \theta^{\mu \nu} {\partial^2 \over \partial x^{\mu} \partial y^{\nu}}}  f(x) g(y).
 \end{equation}
In this paper, we will study $U(N)$ gauge theories that admit a natural noncommutative generalization \cite{Szabo:2001kg,Douglas:2001ba}. 

Noncommutative gauge theories furnish us with an excellent test-case for the questions that we wish to study. These theories are nonlocal but yet relativistic and perturbative. Thus their S-matrices have novel analytic
properties that are easily accessible within perturbation theory. This is exactly what we need.

As we have explained above, nonlocal interactions can lead to an analytic structure that
is very different from that of ordinary quantum field theories. Indeed, noncommutative amplitudes, even at tree level, possess essential singularities in the complex plane. Now, the 
tree level recursion relations that we mentioned above --- the BCFW recursion relations --- are quite delicate. For example, the addition of generic higher order terms in the action makes
them inapplicable. Noncommutative field theories have an infinite number of derivatives 
in their interaction vertices. Naively, this might lead us to think that the BCFW 
recursion relations are not applicable at all. This naive expectation is incorrect.

The BCFW recursion relations are often written down for, what are called, color-ordered amplitudes. A color-ordered amplitude is a part of the full scattering amplitude that is obtained by summing over all double line graphs that have the same cyclic ordering of external momenta (the full amplitude requires us to consider color-ordered amplitudes with all possible different cyclic orderings). Now, color-ordered amplitudes also have a special place in the study of noncommutative field theories. This is because they differ from the corresponding amplitude in ordinary 
theories only by a calculable phase \cite{Filk:1996dm}! This remarkable property allows us to directly apply the BCFW recursion relations to noncommutative amplitudes.

At one-loop we can have both planar and non-planar double line graphs and this makes the situation somewhat more subtle. In ordinary theories, all one-loop amplitudes
can be written in terms of three basic loop integrals --- boxes, triangles and bubbles --- 
apart from rational remainders. The result mentioned above allows us to directly generalize
this decomposition to one-loop noncommutative planar amplitudes.

However, for noncommutative, non-planar amplitudes, 
we require a larger basis of loop integrals (this basis is shown in table \ref{mastertable}). Nevertheless, the principle, that arbitrarily complicated one-loop amplitudes can be reduced to linear combinations of a small set of basis loop integrals with coefficients that are related to products of tree amplitudes, remains true. 

Finally, we go on to study the noncommutative version of the ${\mathcal N}=4$ SYM theory.
In \cite{ArkaniHamed:2008gz}, it was pointed out that, in a certain sense, the ordinary ${\mathcal N}=4$ SYM theory has a very simple one-loop S-matrix because only boxes (and no triangles or bubbles) appear in one-loop amplitudes. 
We find that a very similar property holds for the noncommutative theory. There are no bubble integrals in the decomposition of one-loop non-planar amplitudes. Moreover, while triangles do occur, their coefficients are completely controlled by box coefficients. 

In ordinary gauge theories, one-loop amplitudes contain purely rational pieces that have
no branch cut singularities. These terms are inaccessible to cuts and novel methods 
are required to calculate them \cite{Bern:2005cq,Berger:2006ci}.  The origin of these terms is interesting. A purely four dimensional analysis might lead us to believe that one-loop amplitudes can be reduced to a sum of boxes, triangles and bubbles; repeating this analysis carefully within dimensional regularization reveals a possible rational remainder \cite{Bern:1992em,Bern:1995db}. However, in theories with  good UV properties, such as ${\mathcal N}=4$ SYM, these terms are absent \cite{Bern:1994zx,Bern:1994cg}. In the same way, for noncommutative, non-planar amplitudes, as we show in Appendix \ref{reductionnoncommut}, a four dimensional reduction of one-loop amplitudes gives us the correct answer. Dimensional regularization does not introduce any subtleties in this reduction procedure. Despite this, noncommutative non-planar amplitudes, in non-supersymmetric gauge theories, do have terms that are entirely free of branch cuts. These terms enter the amplitude through tadpole diagrams that do not vanish in the noncommutative theory \cite{Hayakawa:1999zf,Matusis:2000jf}. It is know that these terms do not appear in supersymmetric theories \cite{Matusis:2000jf}. We defer the problem of calculating these terms for non-supersymmetric theories to a future study.

An overview of the rest of the paper is as follows. In section \ref{treelevelsection}, we show 
that the BCFW recursion relations can be used to calculate tree-level amplitudes in noncommutative gauge theories. In section \ref{oneloopsection}, we show that one-loop amplitudes can also be calculated using on-shell methods. In section \ref{n=4section}, we study the S-matrix of ${\mathcal N}=4$ noncommutative SYM theory, at tree and loop-level. The S-matrix of the noncommutative ${\mathcal N}=4$ SYM theory is structurally very simple, just like the S-matrix of the ordinary ${\mathcal N}=4$ SYM theory. In section \ref{examplesection}, we work out some explicit examples of scattering amplitudes to elucidate these ideas. Appendix \ref{reductionnoncommut} contains a proof of some of the claims made in section \ref{oneloopsection}.

\section{On-Shell Methods for Noncommutative Tree Amplitudes}
\label{treelevelsection}
In this section, we show how to generalize the BCFW recursion relations to 
noncommutative gauge theories. 

\subsection{Review}
Consider a scattering amplitude of $n$ particles in a $U(N)$ gauge theory. The amplitude depends on
the momentum, color and polarization of each particle. Since the amplitude is a gauge invariant object, it is possible to write it as a sum over traces. At tree-level we can only get a single trace. Thus, it must be true that the amplitude ${\mathcal A}^{\mathrm t}$ can be written as (see \cite{Bern:1994zx} and references there)
\begin{equation}
\label{colorordered}
{\mathcal A}^{\mathrm t} = \sum_{\pi \in S_n/Z_n} A_{\pi}^{\mathrm t} {\rm Tr}\left(T^{a_{\pi(1)}} \ldots T^{a_{\pi(n)}}\right),
\end{equation}
where $a_i$ indexes the color of particle $i$ and the matrices $T^i$ are the adjoint generators. The sum is over all the set of all permutations modulo cyclic permutations. 
The coefficients of the traces, $A_{\pi}^{\mathrm t}$ are called color-ordered amplitudes. We use the superscript $t$ to indicate that this is a tree amplitude.

It is simple to prove the statement above. We merely reformulate perturbation theory in double line notation. A double line graph has the property that it uniquely specifies the {\it cyclic ordering} of the external momenta.\footnote{In a single-line graph, the notion of cyclic ordering of external momenta is ill-defined.} Hence, the sum over all double line graphs naturally leads to the structure \eqref{colorordered}.  

The full amplitude ${\mathcal A}^{\mathrm t}$  is, of course, invariant under permutations of the different particles but when we speak of a color-ordered amplitude we must fix a particular cyclic ordering. We have suppressed the dependence on the external momenta and helicities in \eqref{colorordered} but, at times below, we will show this explicitly. 

\subsubsection{BCFW extension}

Next, we briefly review spinor helicity variables (see \cite{Dixon:1996wi} and references there). 
Given an on-shell momentum for a massless particle, we can decompose it into spinors using
\begin{equation}
p_{\alpha \dot{\alpha}} = p_{\mu} \sigma^{\mu}_{\alpha \dot{\alpha}} = \l_{\alpha} \lb_{\alpha}.
\end{equation}
We can take dot products of two momenta using
\begin{equation}
2 p_1 \cdot p_2 = \< \l_1, \l_2 \> \left[ \lb_1, \lb_2 \right],
\end{equation}
where
\begin{equation}
\<l_1, l_2\> = \epsilon_{\alpha \beta} \l_1^{\alpha} \l_2^{\beta}, \qquad  \left[l_1, l_2\right] = \epsilon_{\dot{\alpha} \dot{\beta}} \l_1^{\dot{\alpha}} \l_2^{\dot{\beta}}.
\end{equation}
In terms of these spinors, gauge boson polarization vectors can be chosen to be
\begin{equation}
\epsilon^+_{\alpha \dot{\alpha}} = {\mu_{\alpha} \lb_{\dot{\alpha}} \over \< \mu, \l \>},  \qquad 
\epsilon^-_{\alpha \dot{\alpha}} = {\l_{\alpha} \bar{\mu}_{\dot{\alpha}} \over \left[\lb, \bar{\mu}\right]},
\end{equation}
where $\mu,\bar{\mu}$ are arbitrary reference spinors.

Consider an arbitrary color ordered gauge boson amplitude. We denote the helicity of the first gauge boson by $\sigma_1$ and that of the $n^{\rm th}$ gauge boson by $\sigma_n$. Now, deform the momenta and polarization vectors of these particles according to
\begin{equation}
\label{bcfwextension}
\begin{split}
\l_1(z) &= \l_1, \:  \lb_1(z) = \lb_1 + z \lb_n, \:  \l_n(z)=\l_n-z\l_1, \:  \lb_n(z)=\lb_n,  \quad  {\rm if} (\sigma_1, \sigma_n) = (-1,1), \\
\lambda_1(z) &= \lambda_1 + \lambda_n z, \:  \lb_1(z) = \lb_1, \: \l_n(z)=\l_n, \:  \lb_n(z) = \lb_n - z \lb_1, \quad  {\rm otherwise.} 
\end{split}
\end{equation}
Here $z$ is an arbitrary complex number. Note that while $z$ can become large which makes individual components of the momentum associated with particle $1$ and $n$ large, each momentum stays on shell. This is called the BCFW extension. It was shown in \cite{Britto:2004ap,Britto:2005fq} that, under this extension, the amplitude
\begin{equation}
\label{largezbehaviour}
A^{\mathrm t}\left(\{\sigma_1, \l_1(z),\lb_1(z)\} \ldots \{\sigma_n, \l_n(z),\lb_n(z)\}\right)  \xrightarrow[z \rightarrow \infty]{} {\rm O}\left({1 \over z}\right).
\end{equation}
This surprising result allows us to write down recursion relations for tree amplitudes. Tree amplitudes develop simple poles in $z$ whenever an intermediate propagator goes on shell and the residue at each pole is just the product of two smaller tree amplitudes. Since the amplitude dies off at large $z$, we can completely reconstruct it from these poles i.e. from lower point tree amplitudes. This leads to the BCFW recursion relations.
\begin{equation}
\label{bcfwrecursion}
\begin{split}
&A^{\mathrm t}\left(\{\sigma_1, \l_1(z),\lb_1(z)\} \ldots \{\sigma_n, \l_n(z),\lb_n(z)\}\right)\\ &=\sum_{\substack{j=2 \\ \sigma_{\rm int} = \pm 1}}^{n-2} {1 \over \left(p_1(z) + \sum_{i=2}^j p_i\right)^2} \left[A^{\mathrm t}\left(\{\sigma_1,\l_1(z_p^j), \lb_1(z_p^j)\} \ldots \{\sigma_j,\l_j, \lb_j\}, \{\sigma_{\rm int},\l_{\rm int}, \lb_{\rm int}\}\right)\right. \\
&\times \left. A^{\mathrm t}\left(\{-\sigma_{\rm int},\l_{\rm int}, -\lb_{\rm int}\},\{\sigma_{j+1},\l_{j+1}, \lb_{j+1}\} \ldots \{\sigma_{n},\l_{n}(z_p^j), \lb_{n}(z_p^j)\}\right) \right],
\end{split}
\end{equation}
where $z_p^j,p_{\rm int}$ are defined by
\begin{equation}
\left(p_1(z_p^j) + \sum_{i=2}^j p_i\right)^2 = 0,  \qquad  p_{\rm int} = p_1(z_p^j) + \sum_{i=2}^j p_i,
\end{equation}
and the sum over $\sigma_{\rm int}$ runs over all possible intermediate helicities.

In other dimensions, spinor helicity variables are not available (see, though, the recent paper \cite{Cheung:2009dc}) but a very similar set of recursion relations can be derived for tree amplitudes in a gauge theory in any number of dimensions \cite{ArkaniHamed:2008yf}.

\subsection{BCFW Analysis for Noncommutative Amplitudes}
Finally, let us briefly review perturbation theory for noncommutative gauge theories. 
The generalization of the machinery above to noncommutative theories will then become apparent.

Perturbation theory, for noncommutative theories, is most usefully formulated in double line notation. Indeed, this notation is convenient even if we have no gauge fields in the picture. We refer the reader to \cite{Minwalla:1999px} for an analysis of noncommutative perturbative dynamics. 

\begin{figure}[!h]
\begin{center}
\scalebox{0.4}{\input{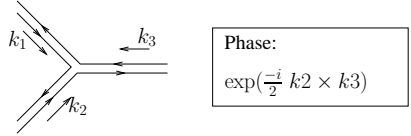}}
\caption{Three Point Vertex}
\label{threept}
\end{center}
\end{figure}

In double line notation,  each noncommutative vertex develops an additional phase factor over the corresponding vertex in the ordinary theory.
\begin{equation}
\label{ordncphase}
\begin{split}
V^{\rm nc}(k_1 \ldots k_n) &= V^{\rm ord}(k_1 \ldots k_n) \exp\left[{-i \over 2} \sum_{i < j} k_i \times k_j\right], \\
k_i \times k_j &= k_i^{\mu} k_j^{\nu} \theta_{\mu \nu}.
\end{split}
\end{equation}
For example, the three point vertex and its associated phase factor are shown in figure \ref{threept}.
Note that we have marked the flow of index lines and in calculating the phase we follow these arrows which causes us to go around the vertex in a counter-clockwise orientation. 

Thus we see that, even tree level noncommutative amplitudes have essential singularities in the complex plane. At first sight, we might worry that these essential singularities will 
complicate amplitudes with all sorts of phases. This is, indeed, true of the full scattering amplitude. In fact, even the full three point amplitude (obtained by summing over the two possible cyclic orderings of external legs) involves $\sin(k_2 \times k_3)$. What makes it possible to use the BCFW recursion relations in noncommutative theories is a simple and remarkable result about {\it color-ordered amplitudes} in noncommutative theories that was first proved by Filk \cite{Filk:1996dm}.

It was shown in \cite{Filk:1996dm} that {\it planar color-ordered amplitudes} in a noncommutative gauge theory can be obtained simply from their commutative counterparts; one simply multiplies the ordinary color ordered amplitude by a momentum dependent phase. To be specific
\begin{equation}
\begin{split}
\label{ordncrelation}
A^{\rm nc}(k_1 \ldots k_n) &= A^{\rm ord}(k_1 \ldots k_n) \phi(k_1 \ldots k_n),\\
\phi(k_1 \ldots k_n) &= \exp\left[{-i \over 2} \sum_{i < j} k_i \times k_j\right].
\end{split}
\end{equation}
We review the proof of this result following \cite{Ishibashi:1999hs,Minwalla:1999px} referring the reader to those papers for more details. 

As we have already explained, noncommutative perturbation theory is most conveniently formulated in double-line notation. 
Now, for a planar graph, we can associate a `momentum' with each index line  \cite{Gross:1982at}.
This `momentum' flows in the direction of the index line. 
The actual momentum flowing through a propagator is the difference 
of the `index-momenta' flowing on the two sides of the propagator,
 $p_{i j} = l_i - l_j$ where $i,j$ are labels for the index lines 
on the two sides of the propagator.

For example, in figure \ref{threept}, we can write 
\begin{equation}
k_1 = l_1 - l_2, \qquad  k_2 = l_2-l_3, \qquad  k_3=l_3-l_1
\end{equation}

The phase associated to a vertex in \eqref{ordncphase} is then
\begin{equation}
\sum_{i < j} k_i \times k_j = \sum_{i = 1}^{n-1} l_i \times l_{i + 1} + l_n \times l_1.
\end{equation}
The advantage of this approach is clear; it allows us to write the total phase factor at each vertex as a sum of phase factors, each of which comes from a propagator. Now, the phase factors for each internal propagator cancel at its two ends. This leaves us only with phase factors from the external legs. From here, it is clear that \eqref{ordncrelation} follows.

However, this is very extremely fortuitous! To obtain the noncommutative planar amplitudes, we just need the color-ordered {\it ordinary} amplitudes. These can be calculated by the standard recursion relations. We then multiply this answer by the appropriate phase to obtain the noncommutative amplitude. The essential singularities in noncommutative planar amplitudes have been controlled!

The result above is very simple once a knowledge of recursion relations and noncommutative perturbation theory is put together. However, this should not obscure the fact that the applicability of recursion relations to noncommutative gauge theories relies on a rather remarkable structure. To emphasize this, consider what happens if we add generic higher order terms --- say a $F^4$ term --- to the action. Recursion relations rely crucially on the property \eqref{largezbehaviour}. It is easy to see, from the the derivation in \cite{ArkaniHamed:2008yf}, that generic higher order terms will spoil this property. In noncommutative theories, the action contains an infinite number of higher derivative terms. In spite of this, noncommutative tree amplitudes are amenable to recursion relations! 

\section{One-Loop Analysis}
\label{oneloopsection}
\subsection{Review}
\label{oneloopordinaryreview}
One loop amplitudes receive contributions from both planar and non-planar double-line graphs. The planar graphs provide a single trace contribution to the amplitude whereas the non-planar graphs lead to a double trace contribution. In analogy with \eqref{colorordered}, we then obtain the following trace decomposition for $U(N)$ gauge theory amplitudes (see \cite{Bern:1991aq} and references there),
\begin{equation}
\label{oneloopcolor}
\begin{split}
&{\mathcal A}^{\mathrm 1 \ell} = N \sum_{\pi \in S_n/Z_n} A_{\pi}^{\mathrm p} {\rm Tr}\left(T^{a_{\pi(1)}} \ldots T^{a_{\pi(n)}}\right) \\
&+ \sum_{j=1}^{[n/2]} \sum_{\pi \in S_n/S_{n;j}}  A^{\mathrm np}_{j;\pi} {\rm Tr}\left(T^{a_{\pi(1)}} \ldots T^{a_{\pi(j)}}\right) {\rm Tr}\left(T^{a_{\pi(j+1)}}  \ldots T^{a_{\pi(n)}}\right).
\end{split}
\end{equation}
Here $S_{n;j}$ is the set of all permutations that cyclically permutes the two sets $(1 \ldots j)$, $(j+1 \ldots n)$ within themselves. Thus, the sum above produces all possible single and double trace structures. The superscript $1 \ell$ indicates a 1-loop amplitude, the superscript ${\mathrm p}$ stands for planar while ${\mathrm np}$ stands for non-planar.
\subsubsection{One Loop Analysis in Ordinary Theories}
The reconstruction of one-loop amplitudes using tree amplitudes has a long history (see \cite{Bern:1991aq,Bern:1993mq,Bern:1994zx,Bern:1994cg,Bern:1995db} and references there). This work has been extended in the past few years after the revival of interest in on-shell techniques \cite{Britto:2004nc,Bern:2005cq,Berger:2006ci,Forde:2007mi,ArkaniHamed:2008gz}. We refer the reader to the review \cite{Bern:2007dw}. Briefly, it is now possible, starting with just the on-shell three point amplitude, to systematically reconstruct the one-loop S-matrix for an ordinary gauge theory. This process also lends itself to easy automation \cite{Berger:2008sj}.

 The construction of S-matrix elements, in ordinary theories,  at one-loop proceeds in two steps. The first is to show 
that all one-loop amplitudes can be written as a sum of 3 basic scalar integrals --- boxes, triangles and bubbles --- with coefficients that are rational functions of the external momenta, plus a purely rational remainder. Next, one looks for efficient methods to reconstruct these coefficients and the rational piece of the amplitude.

There are two ways to understand this integral reduction procedure. The first is to look upon 
this as an elaborate application of partial fractions. In 4 dimensions, it is possible to use partial fractions and Passarino-Veltman reduction \cite{Passarino:1978jh} to reduce an arbitrary one-loop integral to a sum of scalar boxes, triangles and bubbles. However, loop integrals need to be dimensionally regulated and repeating this process in $4+2\epsilon$ dimensions \cite{Bern:1992em} leads to an additional rational remainder. This is the philosophy adopted in \cite{Ossola:2006us} and this is also what we shall use in Appendix \ref{reductionnoncommut}.

However, this reduction procedure can also be understood more physically. One-loop amplitudes have branch cuts; the discontinuity across such a cut can be calculated using the celebrated Cutkowski rules \cite{Cutkosky:1960sp}. However, this discontinuity when considered as an analytic function of the remaining kinematic invariants can, itself, have branch cuts. The discontinuity of the discontinuity is given by putting 3 internal lines on shell. In four dimensions, the maximum number of internal lines that can be put on shell is four. Thus, a sum of boxes, triangles and bubbles is enough to reproduce the most general branch-cut structure at one-loop. When supplemented with a rational remainder, this is enough to reproduce the most general analytic 
structure of an ordinary gauge theory amplitude. 

In the past few years, efficient techniques have been developed to perform this integral 
reduction for an arbitrary one-loop amplitude (see \cite{Forde:2007mi,Berger:2008sj,ArkaniHamed:2008gz} and references there).

In this discussion, we should remember that ordinary massless gauge theories have both UV and IR divergences. This is also true of planar amplitudes in the noncommutative theory. Most of our discussion below will deal with non-planar, noncommutative amplitudes. At one-loop, these amplitudes are UV convergent but have IR divergences. We regulate these divergences by working in $4 + 2 \epsilon$ dimensions.
\subsubsection{Noncommutative Non-Planar Amplitudes}
\label{reviewncnp}
How much of the usual analysis holds for noncommutative theories? For the planar part of the one-loop amplitude this analysis goes through almost unchanged. We calculate the ordinary color-ordered planar amplitude using the methods described above and then multiply this by a phase as in section \ref{treelevelsection}. 

However, the really interesting properties of noncommutative theories are in the non-planar sector. Noncommutative non-planar gauge theory theory amplitudes at one-loop are UV finite and show remarkable properties such as UV-IR mixing \cite{Minwalla:1999px}. Let us briefly review 
how this comes about. 

\begin{figure}[!h]
\begin{center}
\scalebox{0.4}{\input{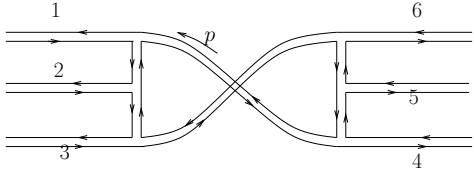}}
\caption{Non-Planar Diagram}
\label{nonplanarexample}
\end{center}
\end{figure}

Consider the non-planar diagram shown in Fig \ref{nonplanarexample}.
What is the phase factor and trace structure associated with this diagram? To write the phase-factor we follow the flow of index lines around the diagram following the convention in figure \ref{threept}. From this it is easy to see that the integrand for the Feynman diagram in figure \ref{nonplanarexample} has a phase factor 
\begin{equation}
\begin{split}
{\rm phase} &= \exp\left\{i p \cdot k \right\} \phi(k_1, k_2, k_3, k_4, k_5, k_6),\\
k_{\mu} &=  {1 \over 2} \theta_{\mu \nu} \left(k_4^{\nu}+k_5^{\nu}+k_6^{\nu} - k_1^{\nu} - k_2^{\nu}- k_3^{\nu}\right), 
\end{split}
\end{equation}
where $\phi$ is defined in \eqref{ordncrelation}, and a trace structure
\begin{equation}
{\rm Trace~Structure} = {\rm Tr}(T^1 T^2 T^3){\rm Tr}(T^4 T^5 T^6).
\end{equation}

It is easy to see that in any non-planar diagram, the trace structure and phase factor are always correlated. The coefficient of the trace 
\begin{equation*}
{\rm Tr}(T^{a_1} \ldots T^{a_j}) {\rm Tr}(T^{a_{j+1}} \ldots T^{a_n}),
\end{equation*}
always comes with a phase factor of
\begin{equation*}
\exp\left[i p \cdot k \right]\phi\left(k_1, \ldots k_n\right)
\end{equation*}
with
\begin{equation}
\label{kdef}
k_{\mu} = {1 \over 2} {\theta_{\mu \nu}}  \left(\sum_{i={j+1}}^n k_i^{\nu}- \sum_{i=1}^{j} k_i^{\nu} \right),
\end{equation}
where $p$ is marked from $j+1 \rightarrow 1$.
We see, thus, that in non-planar graphs the simplest scalar integrals we obtain are of the form\footnote{In this expression and other expressions below, it should be understood that $q_0 = 0$. We do not separate $(p + q_0)^2$ from the other propagators only because this allows for more compact formulae in what follows}
\begin{equation}
\label{prototype}
I = i \int {e^{i p \cdot k}  \over \prod_{i=0}^{r} \left[(p+q_i)^2 + i \epsilon\right]} {d^{4+2\epsilon} p \over (2 \pi)^{4+2\epsilon}}.
\end{equation}
Later we will see that non-planar tensor integrals can be simply related to derivatives
of the integral in \eqref{prototype}. 

The integral in \eqref{prototype} is UV convergent because of the phase factor $e^{i p \cdot k}$. However, if we do introduce a UV cutoff this modifies not only the UV properties of the theory but also its IR properties! This is called UV-IR mixing \cite{Minwalla:1999px}.

Let us analyze these integrals a bit more to study their analytic structure.  We introduce Feynman parameters and Wick rotate to get
\begin{equation}
\label{wickrotated}
I = (-1)^{r}  \int {d^{4+2\epsilon} p_E \over (2 \pi)^{4+2\epsilon}} d x_i  {e^{i p_E \cdot k - \sum x_i (q_i \cdot k)} d x_i\delta\left(1 - \sum x_i \right)\over \left(p_E^2 + \Delta \right)^{r+1} },
\end{equation}
where,
\begin{equation}
\Delta = -\sum_i q_i^2 x_i + \left(\sum_i q_i x_i\right)^2.
\end{equation}
Note that $k$ is purely spatial in a unitary noncommutative theory \cite{Gomis:2000zz,Seiberg:2000gc}, which is what allows us to Wick rotate \eqref{prototype} despite the ${p \cdot k}$ in the exponent.
The integral over the momenta can be done using
\begin{equation}
\label{ncnpbessel}
\begin{split}
I &= (-1)^r \int \exp \left[i p_E \cdot k - \sum x_i (q_i \cdot k) {-\beta \left(p_E^2 + \Delta\right)} \right] {\delta\left(1 - \sum x_i \right) \, d x_i \, d^{4 + 2\epsilon} p_E \, {\beta^r d \beta} \over (2 \pi)^{4+2\epsilon} \Gamma(r+1)}\\
&= {(-1)^{r} |k|^{r-1-\epsilon} \over 2^r(2 \pi)^{2+\epsilon} \Gamma(r+1)} \int dx_i \delta(1 - x_i)  e^{-i \sum x_i (q_i \cdot k)} {K_{r-1-\epsilon}\left(|k| \sqrt{\Delta}\right) \over \left(\sqrt{\Delta}\right)^{r - 1-\epsilon}},
\end{split}
\end{equation}
where $K$ is a modified Bessel function \cite{Abramowitz:1965:HMF} and $|k| = \sqrt{-k^2}$ is the spatial length of $k$.

In ordinary theories, it is possible to explicitly do the integral over Feynman parameters and expand scalar boxes, triangles and bubbles using logarithms
and dilogarithms \cite{'tHooft:1978xw}. We are unaware of a similar expansion for noncommutative integrals.

For large argument, $x >> 1$, the Bessel function goes to $K_{r}(x) \xrightarrow[x >> 1]{} \sqrt{\pi \over 2 x} e^{-x}$. Thus, the integrals \eqref{prototype} also have essential singularities in the complex plane. 

\subsection{A Basis of Master Integrals}

We now discuss how noncommutative non-planar graphs can be reduced to a small set of basis integrals. As we have already mentioned, this basis is larger than the corresponding basis in ordinary theories. 
The integral \eqref{prototype} has branch cut singularities. The position of these singularities is given by the Landau equations \cite{eden1966asm}. Moreover, just as in ordinary theories, we can compute the discontinuity across the branch cut by cutting the corresponding Feynman graph. As in ordinary theories, this discontinuity may itself have a discontinuity which is given by a triple cut; in four dimensions the maximal cut we can make is a four-cut.

However, {\it unlike} ordinary theories, scalar boxes, triangles and bubbles {\it cannot} 
reproduce the most general branch cut structure in noncommutative non-planar amplitudes. 
This is because of the anisotropic phase-factor $e^{i p \cdot k}$ in \eqref{prototype}. In Appendix \ref{reductionnoncommut}, we show how this phase factor forces us to include additional elements in the basis of master integrals. In section \ref{sectionfindingcoeffs}, we show how 
these additional elements can be understood physically from the branch-cut structure of noncommutative non-planar integrals. This is simplest to understand for a four-cut so we refer the interested reader to subsection \ref{subsectionboxcoeffs}.

It is shown in Appendix \ref{reductionnoncommut} that all one-loop non-planar amplitudes in noncommutative gauge theories may be decomposed as
\begin{equation}
\label{decompositionnoncommut}
\begin{split}
A^{\mathrm np}_{j;\pi} &=  \sum_{{\alpha_4}} \int {\sum_{m = 0}^{1} A_{\alpha_{4}}^{(m)} (p \cdot k)^m e^{i p \cdot k} \over \prod_{i = 0}^{3}\left[(p + q^{\alpha_{4}}_i)^2 + i \epsilon\right]} \, {d^{4+2\epsilon} p \over (2 \pi)^{4 + 2\epsilon}} \\
&+ \sum_{\alpha_{3}} \int {\sum_{m = 0}^{3} B_{\alpha_{3}}^{(m)}(p \cdot k)^m  e^{i p \cdot k} \over \prod_{i = 0}^{2}\left[(p + q^{\alpha_{3}}_i)^2+i \epsilon \right]} \, {d^{4+2\epsilon} p \over (2 \pi)^{4 + 2\epsilon}} \\
&+  \int {\sum_{m = 0}^{2} C^{(m)}(p \cdot k)^m  e^{i p \cdot k} \over \prod_{i = 0}^{1}\left[(p + q_i)^2 + i \epsilon\right]} \, {d^{4+2\epsilon} p \over (2 \pi)^{4 + 2\epsilon}}\\
&+ {\mathcal R} + {\rm O}\left(\epsilon\right).
\end{split}
\end{equation}
Here $A,B,C,D$ are rational functions of the external momenta and $\theta$ multiplied by a possible phase that is linear in $\theta$ and bilinear in the external momenta. 
These coefficients are free free of branch-cut singularities. The index $\alpha_{n}$ runs over different partitions of the external momenta into $n$ sets for a given noncommutative non-planar amplitude $A_{\pi}^{\mathrm np}$. 
It serves to remind us that in the expansion of any amplitude, there are {\it several} distinct boxes and triangles. On the other hand, as explained above equation \eqref{kdef}, $k$ is the same for every integral that appears in the expansion \eqref{decompositionnoncommut} above. 
We emphasize that the expansion above is correct as a Laurent series in $\epsilon$ up to terms of ${\rm O}(\epsilon)$. This basis of integrals is shown in table \ref{mastertable}.
\begin{table}
\caption{A Basis for Non-Planar One Loop Amplitudes}
\label{mastertable}
\begin{center}
\begin{tabular}{|c|c|c|}
\hline
Mnemonic & Expression \\ \hline
\includegraphics*[viewport=0 0 200 210,height=0.75cm]{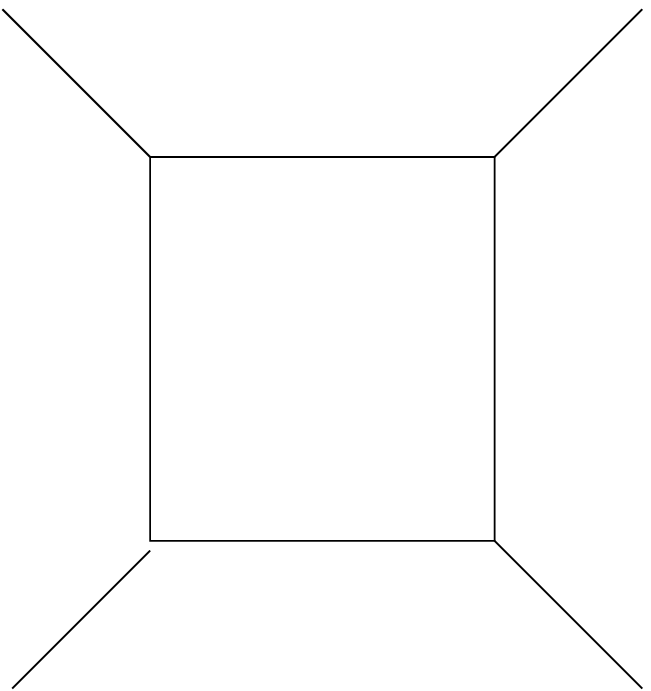}& $\int   {d^{4+2\epsilon} p \over (2 \pi)^{4+2\epsilon}}{e^{i p \cdot k} \over  \left(p^2+i\epsilon\right)\left((p+q_1)^2+i \epsilon\right)\left((p + q_2)^2 + i \epsilon\right) \left((p+q_3)^2 + i \epsilon\right)}$   \\ \hline
\includegraphics*[viewport=0 0 200 210,height=0.75cm]{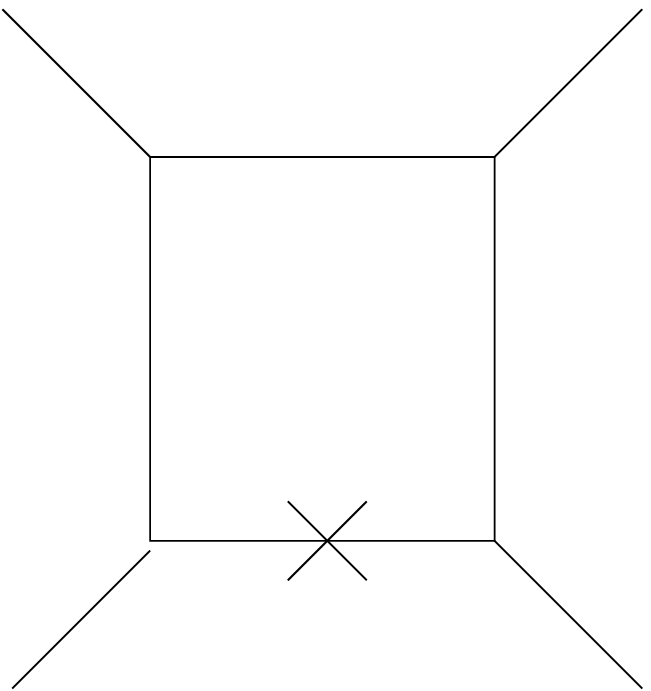}& $\int   {d^{4+2\epsilon} p \over (2 \pi)^{4+2\epsilon}}{(p \cdot k) e^{i p \cdot k} \over  \left(p^2 + i \epsilon\right)\left((p+q_1)^2+i \epsilon\right)\left((p + q_2)^2 + i \epsilon\right) \left((p+q_3)^2 + i \epsilon\right)}$   \\ \hline
\includegraphics*[viewport=0 0 170 100,height=0.75cm]{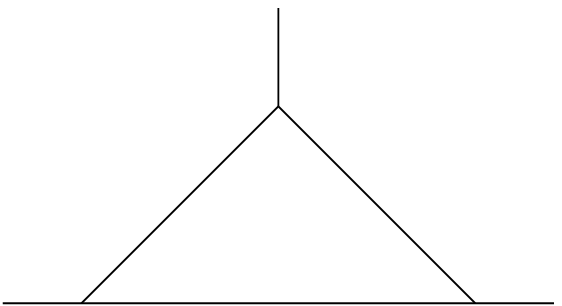} & $\int  {d^{4+2\epsilon} p \over (2 \pi)^{4+2\epsilon}} {e^{i p \cdot k} \over  \left(p^2 + i \epsilon\right)\left((p+q_1)^2+i \epsilon\right)\left((p + q_2)^2 + i \epsilon\right) } $ \\\hline
\includegraphics*[viewport=0 0 170 100,height=0.75cm]{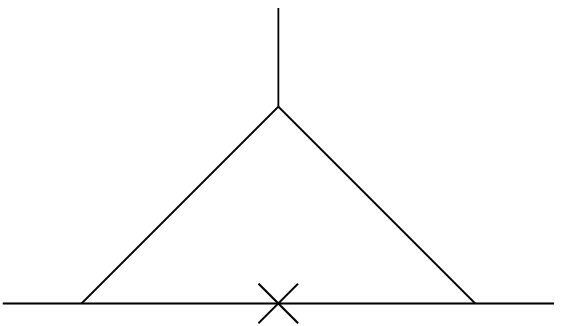} & $\int  {d^{4+2\epsilon} p \over (2 \pi)^{4+2\epsilon}} {(p \cdot k) e^{i p \cdot k} \over  \left(p^2 + i \epsilon\right)\left((p+q_1)^2+i \epsilon\right)\left((p + q_2)^2 + i \epsilon\right) } $ \\\hline
\includegraphics*[viewport=0 0 170 100,height=0.75cm]{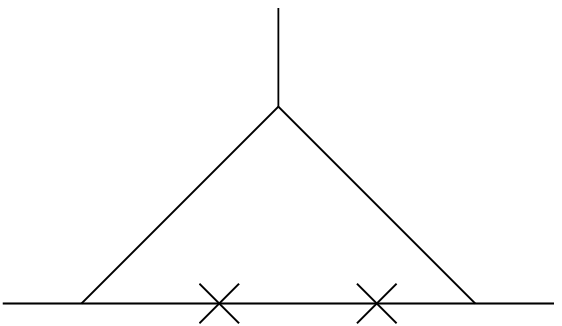} & $\int  {d^{4+2\epsilon} p \over (2 \pi)^{4+2\epsilon}} { (p \cdot k)^2 e^{i p \cdot k} \over  \left(p^2 + i \epsilon\right)\left((p+q_1)^2+i \epsilon\right)\left((p + q_2)^2 + i \epsilon\right) }$ \\\hline
\includegraphics*[viewport=0 0 170 100,height=0.75cm]{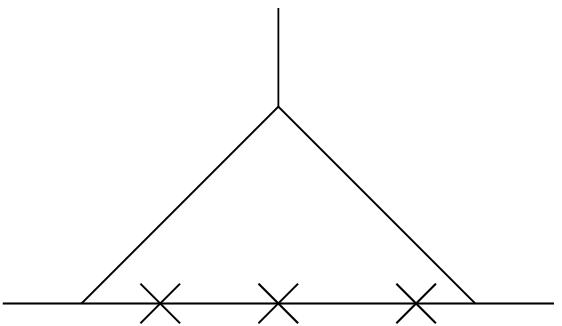} & $\int {d^{4+2\epsilon} p \over (2 \pi)^{4+2\epsilon}} { (p \cdot k)^3 e^{i p \cdot k} \over  \left(p^2 + i \epsilon\right)\left((p+q_1)^2+i \epsilon\right)\left((p + q_2)^2 + i \epsilon\right) }$ \\\hline
\includegraphics*[viewport=0 0 122 32,height=0.6cm]{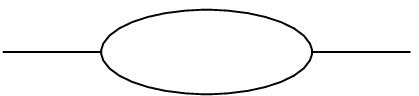} & $\int {d^{4+2\epsilon} p \over (2 \pi)^{4+2\epsilon}} {e^{i p \cdot k}  \over  \left(p^2 + i \epsilon\right)\left((p+q_1)^2+i \epsilon\right) } $\\\hline
\includegraphics*[viewport=0 0 122 36,height=0.6cm]{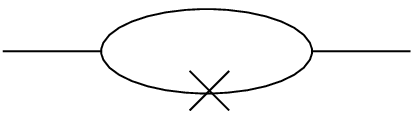} & $\int {d^{4+2\epsilon} p \over (2 \pi)^{4+2\epsilon}} {(p \cdot k) e^{i p \cdot k}  \over  \left(p^2 + i \epsilon\right)\left((p+q_1)^2+i \epsilon\right) }$ \\\hline
\includegraphics*[viewport=0 0 122 36,height=0.6cm]{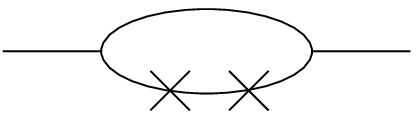} & $\int {d^{4+2\epsilon} p \over (2 \pi)^{4+2\epsilon}} {(p \cdot k)^2 e^{i p \cdot k}  \over  \left(p^2 + i \epsilon\right)\left((p+q_1)^2+i \epsilon\right) }$ \\\hline
1 & Remainder (free of branch-cuts) \\\hline
\end{tabular}
\end{center}
\end{table}

Note that every integral in this basis can be obtained from \eqref{ncnpbessel}. The tensor integrals in table \ref{mastertable} that have insertions of $(p \cdot k)$ in the numerator are related to the scalar integrals by
\begin{equation}
\label{tensorscalarrelation}
\int {\left(p \cdot k\right)^m e^{i p \cdot k}  \over \prod_{i=0}^r \left[(p+q_i)^2 + i \epsilon\right]} {d^{4+2\epsilon} p \over (2 \pi)^{4+2\epsilon}} = |k|^m \left({\partial \over \partial |k|}\right)^m \int { e^{i p \cdot k}  \over \prod_{i=0}^r \left[(p+q_i)^2 + i \epsilon\right]} {d^{4+2\epsilon} p \over (2 \pi)^{4+2\epsilon}},
\end{equation}
where $|k|=\sqrt{-k^2}$. Now we explain how to find the 9 loop coefficients in \eqref{decompositionnoncommut}. 

\subsection{Obtaining the Coefficients}
\label{sectionfindingcoeffs}

\subsubsection{Box Coefficients}
\label{subsectionboxcoeffs}
The Box coefficients are the simplest to find. We will be very explicit for this case.
 First, let us understand the index $\alpha_4$ in \eqref{decompositionnoncommut}. The term $A^{\mathrm np}_{j;\pi}$ in \eqref{oneloopcolor} receives contributions from different boxes, each of which is labeled by a pair of integers $0< i_1 < j, 0 < i_2 < n-j$. The index $\alpha_4$ is shorthand for these two integers. Given these two integers and $j$ we can divide the external momenta into 4 sets
\[
\{k_{\pi_{1}} \ldots k_{\pi_{i_1}}\}, \{k_{\pi_{i_1 + 1}} \ldots k_{\pi_{j}}\}, \{k_{\pi_{j + 1}} \ldots k_{\pi_{j + i_2}}\}, \{k_{\pi_{j + i_2 + 1 }} \ldots k_{\pi_{n}}\}.
\]
For any such partition, we have
\begin{equation}
q_1 = \sum_{m=1}^{i_1} k_{\pi_{m}},  \qquad  q_2 = q_1 + \sum_{m=i_1+1}^{j} k_{\pi_{m}}, 
 \qquad  q_3 = q_2 + \sum_{m=j+i_2+1}^{n} k_{\pi_{m}}.
\end{equation}
When we put 4 lines on shell, we need to solve the equations
\begin{equation}
\label{fouronshell}
p^2 = (p + q_1)^2 = (p + q_2)^2 = (p + q_3)^2 = 0.
\end{equation}
Evidently, this fixes the internal momentum $p$ 
to be one of two possible values \cite{Britto:2004nc}. Let us call these two solutions $p^{\pm}$. For each of
these solutions, we calculate the four-cut
\begin{equation}
\label{fourcutexplicit}
\begin{split}
&{\mathcal C}^{\pm}_{\alpha_4} e^{i p^{\pm} \cdot k} = \sum_{\sigma_{\mathrm int}^i=\pm 1} \left[ A^{\mathrm t}\left(\{\sigma_{\rm int}^{1},p^{\pm}\},\{\sigma^{\pi_{1}},k_{\pi_{1}}\}, \ldots \{\sigma^{\pi_{i_1}},k_{\pi_{i_1}}\},\{-\sigma_{\rm int}^{2},-p^{\pm}-q_1\}\right) \right. \\
&\times A^{\mathrm t}\left(\{\sigma_{\rm int}^{2},p^{\pm}+q_1\},\{\sigma^{\pi_{i_1 + 1}},k_{\pi_{i_1 + 1}}\}, \ldots \{\sigma^{\pi_{j}},k_{\pi_{j}}\},\{-\sigma_{\rm int}^{3},-p^{\pm}-q_2\}\right)\\
&\times A^{\mathrm t}\left(\{\sigma_{\rm int}^{3},p^{\pm}+q_2\},\{-\sigma_{\rm int}^{4},-p^{\pm}-q_3\},\{\sigma^{\pi_{j + i_2+1}},k_{\pi_{j + i_2+1}}\}, \ldots \{\sigma^{\pi_{n}},k_{\pi_{n}}\}\right) \\
&\times \left. A^{\mathrm t}\left(\{\{\sigma_{\rm int}^{4},p^{\pm}+q_3\},\{-\sigma_{\rm int}^{1},-p^{\pm}\}, \{\sigma^{\pi_{j+1}},k_{\pi_{j+1}}\}, \ldots \{\sigma^{\pi_{j + i_2}},k_{\pi_{j + i_2 }}\}\right) \right].
\end{split}
\end{equation}
We wish to emphasize two points here
\begin{enumerate}
\item
The order of particles in the tree amplitudes (up to cyclic permutations) is important. Note that the momenta
appear in a different order in the third and fourth tree amplitudes in \eqref{fourcutexplicit} than they do in the first and second.
\item
It is easy to check that the product of the four tree momenta in \eqref{fourcutexplicit} produces a $p$ dependent phase that we have explicitly extracted in the definition of ${\mathcal C}$.
\end{enumerate}
We need to reproduce this cut, using the terms in \eqref{decompositionnoncommut}. This is done
by solving the two equations
\begin{equation}
\label{boxequations}
\begin{split}
{\mathcal C}_{\alpha_4}^{+} &= A_{\alpha_4}^{(0)} + A^{(1)}_{\alpha_4} \left(p^{+} \cdot k\right) \\
{\mathcal C}_{\alpha_4}^{-} &= A_{\alpha_4}^{(0)} + A^{(1)}_{\alpha_4} \left(p^{-} \cdot k\right) 
\end{split}
\end{equation}
With this definition, $A^{(0)}_{\alpha_4}, A^{(1)}_{\alpha_4}$ become rational functions of the 
external momenta and $\theta$ with a phase that is bilinear in the external momenta and linear in $\theta$. 

We wish to emphasize the contrast with ordinary theories.
In ordinary theories, for each partition, we have only a single box coefficient as opposed to the two coefficients $A^{(0)}$ and $A^{(1)}$ that we have here. There, we add the contribution from both 
solutions of \eqref{fouronshell}, and set the single box coefficient to the sum. In noncommutative gauge theories, this procedure would not correctly reproduce the 4-cut because of the additional p dependent phase in \eqref{fourcutexplicit}.  We need the two different boxes shown in table \ref{mastertable} to accurately reproduce this behavior.

Hence, in ordinary gauge theories, the cuts provide us with more information than we use.  Noncommutative gauge theories, on the other hand, use all the information that is provided by the cuts!

\subsubsection{Triangle Coefficients}
\label{findtriangle}
Now, let us consider the triangle coefficients. Once again, several triangles contribute
to any particular noncommutative non-planar amplitude. Given either an integer $0 < i_1 < j$ {\it or} 
an integer $0 < i_2 < n-j$, we can partition the momenta into three sets.
\begin{equation*}
\begin{split}
&\{k_{\pi_{1}} \ldots k_{\pi_{i_1}}\}, \{k_{\pi_{i_1 + 1}} \ldots k_{\pi_{j}}\}, \{k_{\pi_{j + 1}} \ldots k_{\pi_n}\}  \qquad  {\rm or}\\
&\{k_{\pi_{1}} \ldots k_{\pi_{j}}\}, \{k_{\pi_{j + 1}} \ldots k_{\pi_{j+i_2}}\}, \{k_{\pi_{j + i_2+1}} \ldots k_{\pi_n}\}.
\end{split}
\end{equation*}

Consider the first case (the generalization to the second case is obvious). Here
\begin{equation}
q_1 = \sum_{i=1}^{i_1} k_{\pi_i},  \qquad  q_2 = q_1 +  \sum_{i=i_1+1}^{j} k_{\pi_i}.
\end{equation}
The 3-cut does not freeze the internal momenta; instead it leaves us with one complex parameter. We fix this parameter by solving the equations 
\begin{equation}
\label{threeonshell}
p^2 = (p + q_1)^2 = (p + q_2)^2 = 0, \:  p \cdot k = z.
\end{equation}
Once again, we have {\it two} solutions to these equations, that we will call $p^{\pm}$.

We calculate the three-cut
\begin{equation}
\label{threecutexplicit}
\begin{split}
{\mathcal C}^{\pm}_{\alpha_3} e^{i z} &= \sum_{\sigma_{i}=\pm 1} \left[ A^{\mathrm t}\left(\{\sigma_{\rm int}^{1},p^{\pm}\},\{\sigma^{\pi_{1}},k_{\pi_{1}}\} \ldots \{\sigma^{\pi_{i_1}},k_{\pi_{i_1}}\},\{-\sigma_{\rm int}^{2},-p^{\pm}-q_1\}\right) \right. \\
& A^{\mathrm t}\left(\{\sigma_{\rm int}^{2},p^{\pm}+q_1\},\{\sigma^{\pi_{i_1 + 1}},k_{\pi_{i_1 + 1}}\} \ldots \{\sigma^{\pi_{j}},k_{\pi_{j}}\},\{-\sigma_{\rm int}^{3},-p^{\pm}-q_2\}\right)\\
&\left. A^{\mathrm t}\left(\{\{\sigma_{\rm int}^{3},p^{\pm}+q_2\},\{-\sigma_{\rm int}^{1},-p^{\pm}\},\{\sigma^{\pi_{j  + 1 }},k_{\pi_{j  + 1 }}\} \ldots \{\sigma^{\pi_{n}},k_{\pi_{n}}\}\right) \right].
\end{split}
\end{equation}
Now, the boxes present in the amplitude also contribute to the three cut. Since, we have already calculated the box coefficients, we can write down this contribution:
\begin{equation}
\label{remainderboxes}
R_{\alpha_3}^{\pm} = \lim_{p \rightarrow p^{\pm}} \left[p^2 (p + q_1^{\alpha_3})^2 (p + q_2^{\alpha_3})^2 \sum_{{\alpha_4}}  {\sum_{m = 0}^{1} A_{\alpha_{4}}^{(m)} (p \cdot k)^m  \over \prod_{i = 0}^{3}(p + q^{\alpha_{4}}_i)^2}\right].
\end{equation}

It is easy to see from the general analysis of the growth of tree amplitudes for large BCFW deformations in \cite{ArkaniHamed:2008yf} that ${\mathcal C}^{\pm}_{\alpha_3}$ can grow like $z^3$ at large $z$. However, in contrast to ordinary theories, the remainder $R_{\alpha_3}^{\pm}$ continues to provide a non-vanishing, ${\rm O}\left(1\right)$ contribution for large $z$. This contribution comes from the crossed box diagram on the second line of table \ref{mastertable} corresponding to an integral that has an insertion of $p \cdot k$ in the numerator. 

In ordinary theories, it is possible to project out the contribution from the boxes just by taking the large $z$ limit of $C^{\pm}_{\alpha_3}$ \cite{Forde:2007mi,ArkaniHamed:2008gz}. This is not enough for noncommutative theories. To isolate the triangle coefficients, we need to explicitly subtract off the box-contribution to the three-cut. This leads us to consider the difference,
\begin{equation}
\label{cprimedeftriangle}
{\mathcal C}'_{\alpha_3} = {1 \over 2} \sum_{\pm}\left( {\mathcal C}^{\pm}_{\alpha_3} - R_{\alpha_3}^{\pm}\right).
\end{equation}

In general, both $C^{\pm}_{\alpha_3}$ and $R^{\pm}_{\alpha_3}$ contain terms that die off as ${\rm O}\left({1 \over z}\right)$ for large $z$. However, 
the analysis in Appendix \ref{reductionnoncommut} tells us that these terms must cancel in \eqref{cprimedeftriangle} --- ${\mathcal C'}_{\alpha_3}$ is a {\it polynomial} of degree $3$ in $z$! The triangle coefficients are then just
\begin{equation}
\label{threecutexpansion}
{\mathcal C}'_{\alpha_3} = B^{(0)}_{\alpha_3} + B^{(1)}_{\alpha_3} z +  B^{(2)}_{\alpha_3} z^2 + B^{(3)}_{\alpha_3} z^3 .
\end{equation}
We emphasize that $R^{\pm}_{\alpha_3}$ makes a physical contribution to $B^{(0)}_{\alpha_3}$.

Note, that we could also extract these coefficients directly from the difference in the large $z$ behavior of ${\mathcal C}_{\alpha_3}$ and $R_{\alpha_3}$; at times, the calculation of the three-cut \eqref{threecutexplicit} may simplify in this limit. However, we emphasize that \eqref{threecutexpansion} is true for all $z$, not just large $z$.
In fact, as we remark below, the fact that  ${\mathcal C'}_{\alpha_3}$ is a {\it polynomial} gives us a computationally efficient method of obtaining triangle coefficients even in ordinary gauge theories. This is similar in spirit to the procedure outlined in \cite{Ossola:2006us,Ossola:2007ax}.
\subsubsection{Bubble Coefficients}
\label{subsectionbubblecoefficients}
A single bubble diagram contributes to each noncommutative non-planar amplitude, with $q_1 = \sum_{i=1}^j k_{\pi_i}$. Now, putting two lines on 
shell leaves us with two free parameters. We introduce auxiliary four-vectors $w_1, w_2$, that satisfy $w_i \cdot k = w_i \cdot q_1 = 0, \: w_i \cdot w_j = \delta_{i j}$. In general, individual components of the $w_i$ will take complex values. Now, set 
\begin{equation}
\label{bubblecut}
p^2 = (p + q_1)^2 = 0,  \quad  p \cdot k = z,  \quad  p \cdot w_1 = \cos \theta, \quad  p \cdot w_2 = \sin \theta.
\end{equation}
 We will use $\omega = e^{i \theta}$ and denote the solutions to \eqref{bubblecut} by $p(\omega)$.
As in the previous subsections, we calculate 
\begin{equation}
\begin{split}
{\mathcal C}_{2}(\omega) e^{i z} &= \sum_{\sigma_{i}=\pm 1} \left[ A^{\mathrm t}\left(\{\sigma_{\rm int}^{1},p(\omega)\},\{\sigma^{\pi_{1}},k_{\pi_{1}}\} \ldots \{\sigma^{\pi_{j}},k_{\pi_{j}}\},\{-\sigma_{\rm int}^{2},-p(\omega)-q_1\}\right) \right. \\
&\left. A^{\mathrm t}\left(\{\{\sigma_{\rm int}^{2},p(\omega)+q_1\},\{-\sigma_{\rm int}^{1},-p(\omega)\},\{\sigma^{\pi_{j  + 1 }},k_{\pi_{j  + 1 }}\} \ldots \{\sigma^{\pi_{n}},k_{\pi_{n}}\}\right) \right],
\end{split}
\end{equation}
and the contribution from the higher order terms
\begin{equation}
\begin{split}
R_2(\omega) = \lim_{p \rightarrow p(\omega)} p^2 (p + q_1)^2 &\left[ \sum_{{\alpha_4}}  {\sum_{m = 0}^{1} A_{\alpha_{4}}^{(m)} (p \cdot k)^m  \over \prod_{i = 0}^{3}(p + q^{\alpha_{4}}_i)^2} \right. \\
&\left. + \sum_{\alpha_{3}} {\sum_{m = 0}^{3} B_{\alpha_{3}}^{(m)}(p \cdot k)^m  \over \prod_{i = 0}^{2}(p + q^{\alpha_{3}}_i)^2}\right].
\end{split}
\end{equation}
The object we are interested in is
\begin{equation}
\label{cprimebubble}
{\mathcal C'}_2(\omega) = {\mathcal C}_2(\omega) - R_2(\omega).
\end{equation}
At any given value of the $\omega$, as discussed in Appendix \ref{reductionnoncommut}, several `spurious' terms contribute to \eqref{cprimebubble}. An analysis of \eqref{reduce} in Appendix \ref{reductionnoncommut} tells us that to project out these spurious terms we should adopt the prescription
\begin{equation}
\label{bubblepolynomial}
 {1 \over 4 \pi i} \oint_{\omega=\infty} {d \omega \over \omega} \left[ {\mathcal C}'_{L}(\omega) + {\mathcal C}'_{L}({1 \over \omega})\right] = \sum_{m=0}^{2} C^{(m)} z^m,
\end{equation}
where the contour integral is taken about $\omega = \infty$.  
\subsubsection{The Remainder}
It is shown in Appendix \ref{reductionnoncommut} that terms that are free of branch-cuts do occur in noncommutative non-planar amplitudes. In contrast with ordinary theories, these terms are not strictly rational, because of the possibility of phases that are bilinear in the external momenta.  

In ordinary theories, a rational remainder arises in one loop integral decomposition, because the cuts that are used to obtain the coefficients of boxes, triangles and bubbles are performed in $4$ dimensions whereas the amplitude itself needs to be dimensionally regulated. Dimensional regularization does not cause any difficulties in the integral decomposition of noncommutative non-planar amplitudes that we have outlined above. This is directly linked to the excellent UV properties of these amplitudes. In ordinary theories with good UV behavior --- like ${\mathcal N}=4$ SYM --- rational remainders do not occur either.

In noncommutative non-planar amplitudes, a remainder that is free of branch cuts comes from tadpole diagrams that vanish in the ordinary theory but remain finite in the noncommutative case. 
A modification of the techniques used in ordinary gauge theories \cite{Bern:2005cq,Berger:2006ci} might help in calculating such terms. However, we should note that the physical input used in those studies was that amplitudes should factorize correctly when the sum of a subset of external momenta goes on shell \cite{Bern:1995ix}. The ``rational'' terms here are not required for any such property. Since they are free of branch-cuts, they are not required by unitarity either. In supersymmetric theories, such terms do not occur at all \cite{Hayakawa:1999zf,Matusis:2000jf}.
We defer the study of these non-supersymmetric remainders to a future paper.
\subsection{Application to Ordinary Theories}
\label{ordinaryonelooporientation}
The technique of extracting one-loop integral coefficients that we have outlined above is also applicable to {\it ordinary} gauge theories. Of course, in ordinary gauge theories, where the loop-integrals have no $e^{i k \cdot p}$ factor, the crossed figures in table \ref{mastertable} can all be reduced to ordinary integrals (with no insertions of $p \cdot k$ in the numerator). 
Nevertheless, it is efficient to calculate the coefficients in table \ref{mastertable} in an intermediate step. 
After calculating these coefficients, we drop the $e^{i p \cdot k}$ factor and reduce the tensor integrals to scalar boxes, triangles and bubbles using Passarino-Veltman reduction.\footnote{We emphasize that because of UV-IR mixing, this is {\it not} the same as taking the $\theta \rightarrow 0$ limit of the integrals in table \ref{mastertable}.}

The reason for this intermediate step is that we have an efficient technique to extract the coefficients in table \ref{mastertable}. This is because \eqref{cprimedeftriangle} and \eqref{bubblepolynomial} give us polynomials in $z$. 
For triangle coefficients, we only need to fit a third order polynomial in $z$. On a computer, this is significantly easier (and requires 4 function calls) than extracting the large $z$ behavior of the three-cut.
For bubble coefficients, once again we only need to fit a second order polynomial which removes the need to do some of the integrals outlined in \cite{Berger:2008sj}. 

This is similar to the procedure used in \cite{Ossola:2007ax} except for one important difference. In \cite{Ossola:2007ax}, the starting point is an explicit expression for the integrand of a one-loop amplitude which is then reduced to a sum of boxes, triangles and bubbles. The procedure of unitarity cuts that we have discussed is not enough to reconstruct the entire Feynman integrand for an amplitude. 
Cuts
do give us the `physical' terms in the integrand i.e. the coefficients of boxes, triangles and bubbles. However, as 
the analysis in Appendix \ref{reductionnoncommut} shows, 
Feynman diagrams also give us `spurious' terms  that integrate to zero. These terms are not physical (they even depend on the choice of gauge) and, in general, we should not expect the product of on-shell tree amplitudes to reproduce them faithfully.
\section{${\mathcal N}=4$ Noncommutative SYM}
\label{n=4section}
We now discuss the S-matrix of noncommutative the $U(N), {\mathcal N}=4$ SYM theory. We will find that, as in the ordinary case, the S-matrix of noncommutative ${\mathcal N}=4$ SYM is structurally simple. In this section, we start by reviewing the on-shell techniques for ordinary ${\mathcal N}=4$ SYM developed in \cite{ArkaniHamed:2008gz}. We will then extend our analysis of the non-planar sector of noncommutative pure gauge theories from the previous section to noncommutative ${\mathcal N}=4$ SYM.

\subsection{Review}
First, we discuss a convenient parameterization of the particle content of the ${\mathcal N}=4$ theory using what is called `on-shell superspace' \cite{Nair:1988bq,Mandelstam:1982cb,Ferber:1977qx}. Our treatment here, closely follows \cite{ArkaniHamed:2008gz}.

The idea is to represent on-shell states in the ${\mathcal N}=4$ theory using a Grassmann vector $\eta^I$,
\begin{equation}
\label{etadef}
|\eta, \l, \lb \rangle = e^{ Q_{I \alpha} \eta^{I} w^{\alpha}} |-1, \l, \lb \rangle.
\end{equation}
Here,  $|-1, \l, \lb \rangle$ is the gauge boson state with negative helicity and momentum $\l_{\alpha} \lb_{\dot{\alpha}}$. $Q_{I \alpha}$ are supersymmetry generators in the ${\mathcal N}=4$ theory that transform in the fundamental representation of the R-symmetry $SU(4)$ (indicated by the index $I$) and as right handed Weyl spinors (indicated by the index $\alpha$). The spinor $w$ must satisfy the condition $\langle w, \l \rangle = 1$. $\eta$ has 4 components that transform in the anti-fundamental of $SU(4)$. This notation compactly packages {\it the entire} ${\mathcal N}=4$ multiplet.

With this notation, scattering amplitudes become smooth functions of the $\eta$'s (one $\eta$ for each particle). A negative helicity gauge boson has $\eta = 0$. To obtain information about a positive helicity gauge boson, we need to integrate over $\eta$
\begin{equation}
A(+1, \ldots) = \int A(\eta, \ldots) d^4 \eta
\end{equation}

Under a supersymmetry transformation
\begin{equation}
\label{susyeta}
e^{Q_{I \alpha} \zeta^{I \alpha}} |\e, \l, \lb \> = |\e^I + \<\zeta^I,\l\>,\l,\lb\>
\end{equation}

It was shown in \cite{ArkaniHamed:2008gz} that if we consider tree level scattering in the ${\mathcal N}=4$ theory then under the modified BCFW extension
\begin{equation}
\label{symbcfw}
A^{\mathrm t}(\{\e_1 + z \e_2, \l_1 + z \l_2, \lb_1\},\{\e_2, \l_2, \lb_2 - z \lb_1\}, \ldots) \xrightarrow[z \rightarrow \infty]{} {\rm O}\left({1 \over z}\right).
\end{equation}
Note that here, we need to extend $\eta_1$ in addition to extending the momenta and polarization vectors of particle $1$ and particle $2$.

The proof of this claim is simple. We do a supersymmetry transformation on the left hand side of \eqref{symbcfw} with the parameter
\begin{equation}
\zeta^I_{\alpha} = {\e_1^I \left(\l_2\right)_{\alpha} - \e_2^I \left(\l_1\right)_{\alpha} \over \<\l_1 \l_2\>}
\end{equation}
Using \eqref{susyeta}, we see
\begin{equation}
\label{alltonegative}
\begin{split}
&e^{ \< Q_I, \zeta^I \>} |\{\e_1 + z \e_2, \l_1 + z \l_2, \lb_1\}, \{\e_2, \l_2, \lb_2 - z \lb_1\}, \ldots \> \\ &= |\{0, \l_1 + z \l_2, \lb_1 \},\{0, \l_2, \lb_2 - z \lb_1\}, \ldots \>
\end{split}
\end{equation}

Note that the other $\eta$'s in the state in \eqref{alltonegative} do change by ${\rm O}(1)$. This, of course, cannot affect the large $z$ scaling of the amplitude in \eqref{symbcfw}. Thus, at large $z$, the scattering amplitude in \eqref{symbcfw} has the same scaling as the amplitude of two BCFW extended negative helicity gauge bosons in ${\mathcal N}=4$ SYM theory. However, this was shown to vanish as ${\rm O}\left({1 \over z}\right)$ in \cite{Cheung:2008dn}, which proves our result. 

\subsection{Tree Level Scattering}
Our analysis of tree level scattering in ${\mathcal N}=4$ SYM is parallel to our analysis of pure gauge theories in  section \ref{treelevelsection}.
At tree-level, a scattering amplitude for $n$ particles in a $U(N)$, ${\mathcal N}=4$ noncommutative gauge theory can be decomposed into traces using \eqref{colorordered}.

Moreover, just as in pure gauge theories, in the ${\mathcal N}=4$ SYM theory, planar amplitudes are related to amplitudes in the ordinary theory by \eqref{ordncrelation}.
Now, color-ordered amplitudes in the ordinary ${\mathcal N}=4$ theory can be calculated using the recursion relations given in \cite{ArkaniHamed:2008gz}. We can then use \eqref{ordncrelation} to evaluate color-ordered amplitudes in the noncommutative theory. 

Note that our story relies crucially on ${\mathcal N} = 4$ supersymmetry being maintained. In $\beta$-deformed SYM, for example, different fields have different noncommutativity parameters \cite{Lunin:2005jy}. For these theories, it is not possible to relate all scattering amplitudes to the scattering of negative helicity gauge bosons. Hence, these amplitudes cannot be constructed using the BCFW recursion relations.

\subsection{One Loop Scattering}

Planar one-loop amplitudes in noncommutative ${\mathcal N}=4$ SYM are related to planar one-loop amplitudes in the ordinary theory by a phase given in  \eqref{ordncrelation}. The on-shell techniques that we outlined in subsection \ref{oneloopordinaryreview} apply to any theory and so, can be used to calculate one-loop amplitudes in the ordinary ${\mathcal N}=4$ theory. Once we have this answer, we multiply it by a phase to obtain the noncommutative answer.

It was shown in \cite{BjerrumBohr:2006yw,ArkaniHamed:2008gz} that one-loop amplitudes in the ordinary ${\mathcal N}=4$ theory are structurally very simple. Recall that, at one loop, amplitudes in an ordinary theory (and planar amplitudes in a noncommutative theory) can be written in terms of boxes, triangles, bubbles and a rational remainder. The ``no-triangle hypothesis'' \cite{Bern:2005bb} states that the expansion of ${\mathcal N}=4$ amplitudes only contains boxes. This structural simplicity carries over directly to the planar sector of the noncommutative theory.

Here, we are more interested in the non-planar sector of the theory. Even for this sector, we will see that this simplicity remains, albeit in an altered form.

Non-planar amplitudes in the noncommutative ${\mathcal N}=4$ theory can, in principle be decomposed using \eqref{decompositionnoncommut}. However, we will show below that of the 9 loop coefficients in that equation, only the two box coefficients are independent. The integral expansion of non-planar amplitudes in the noncommutative ${\mathcal N}=4$ theory does not contain any bubbles or rational terms; it does contain triangles but 
the coefficients of these triangles are {\it completely} controlled by the box-coefficients. 

Non-planar amplitudes in the ${\mathcal N}=4$ noncommutative theory have the decomposition
\begin{equation}
\label{decompositionnoncommut2}
\begin{split}
A^{\mathrm{np,sym}}_{j;\pi} &=  \sum_{{\alpha_4}} \int {\sum_{m = 0}^{1} A_{\alpha_{4}}^{(m)} (p \cdot k)^m  e^{i p \cdot k} \over \prod_{i = 0}^{3}\left[(p + q^{\alpha_{4}}_i)^2+ i \epsilon\right]} \, {d^{4+2\epsilon} p \over (2 \pi)^{4 + 2 \epsilon}} \\
&+ \sum_{\alpha_{3}} \int {B_{\alpha_{3}}^{(0)}  e^{i p \cdot k} \over \prod_{i = 0}^{2}\left[(p + q^{\alpha_{3}}_i)^2+i\epsilon\right]} \, {d^{4+2\epsilon} p \over (2 \pi)^{4 + 2 \epsilon}} \\
\end{split}
\end{equation}
As we have explained above, the only independent coefficients in this expression are $A_{\alpha_4}^{(0)}$ and $A_{\alpha_4}^{(1)}$. The single non-zero triangle coefficient $B_{\alpha_3}^{(0)}$ is completely controlled by the box coefficients. We will show that the coefficients $C^{(m)}$ are all zero while the rational remainder ${\mathcal R}$ is already known to be zero in this theory \cite{Matusis:2000jf}.

The box coefficients are calculated using the procedure we outlined in subsection \ref{subsectionboxcoeffs}. Given a partition of the external momenta, we put the internal momenta on shell by demanding \eqref{fouronshell} and  calculate a four-cut
\begin{equation}
\label{fourcutexplicitsym}
\begin{split}
&{\mathcal C}^{\pm}_{\alpha_4} e^{i p^{\pm} \cdot k} = \int \prod d^4 \eta^{ij} \left[ A^{\mathrm t}\left(\{\eta^{31},p^{\pm}\},\{\eta^{\pi_{1}},k_{\pi_{1}}\} \ldots \{\eta^{\pi_{i_1}},k_{\pi_{i_1}}\},\{\eta^{12},-p^{\pm}-q_1\}\right) \right. \\
&\times A^{\mathrm t} \left(\{\eta^{12},p^{\pm}+q_1\}\},\{\eta^{\pi_{i_1 + 1}},k_{\pi_{i_1 + 1}}\}
\ldots \{\eta^{\pi_{j}},k_{\pi_{j}}\},\{\eta^{24},-p^{\pm}-q_2\}\right)\\
&\times A^{\mathrm t}\left(\{\eta^{24},p^{\pm}+q_2\},\{\eta^{43},-p^{\pm}-q_3\},\{\eta^{\pi_{j + i_2+1}},k_{\pi_{j + i_2+1}}\} \ldots \{\eta^{\pi_{n}},k_{\pi_{n}}\}\right) \\
&\times \left. A^{\mathrm t}\left(\{\{\eta^{43},p^{\pm}+q_3\},\{\eta^{31},-p^{\pm}\},\{\eta^{\pi_{j + 1}},k_{\pi_{j + 1 }}\} \ldots \{\eta^{\pi_{j+i_2}},k_{\pi_{j+i_2}}\}\right) \right]
\end{split}
\end{equation}
The only difference with \eqref{fourcutexplicit} is that the sum over the two gluon helicities there, is replaced by an integral over the intermediate $\eta$'s which automatically sums over 
the entire ${\mathcal N}=4$ multiplet. 
Once we have calculated the four-cut, the box coefficients can be calculated using \eqref{boxequations}.

We now turn to the triangle coefficients. Given a partition of the external momenta we put the internal momenta on shell by demanding \eqref{threeonshell}.  We will show that the three-cut ${\cal C}_{\alpha_3}^{\pm}$ vanishes as ${\rm O}\left({1 \over z}\right)$ for large $z$. The box coefficients contribute to ${\mathcal C'}_{\alpha_3}$ (that we defined in \eqref{cprimedeftriangle}) through $R_{\alpha_3}$. Thus only one triangle coefficient --- $B^{(0)}_{\alpha_3}$ --- is non-zero and it is completely determined by the box coefficients.

The proof of this assertion closely follows the proof of the no-triangle hypothesis in \cite{ArkaniHamed:2008gz}. Recall, our procedure for calculating triangle coefficients explained in section \ref{findtriangle}. For any amplitude we choose a partition as explained in subsection \ref{findtriangle} and a set of intermediate momenta satisfying \eqref{threeonshell}. Then we calculate the 3-cut,
\begin{equation}
\label{threecut}
\begin{split}
&{\mathcal C}_{\alpha_3}^{\pm} e^{i z} =  \int \prod d^4 \eta^{ij} \left[ A^{\mathrm t}\left(\{\eta^{31},p^{\pm}\},\{\eta^{\pi_{1}},k_{\pi_{1}}\} \ldots \{\eta^{\pi_{i_1}},k_{\pi_{i_1}}\},\{\eta^{12},-p^{\pm}-q_1\}\right) \right. \\
&\times A^{\mathrm t} \left(\{\eta^{12},p^{\pm}+q_1\}\},\{\eta^{\pi_{i_1 + 1}},k_{\pi_{i_1 + 1}}\}
\ldots \{\eta^{\pi_{j}},k_{\pi_{j}}\},\{\eta^{24},-p^{\pm}-q_2\}\right)\\
&\times \left. A^{\mathrm t}\left(\{\eta^{24},p^{\pm}+q_2\},\{\eta^{31},-p^{\pm}\},\{\eta^{\pi_{j + 1}},k_{\pi_{j +1}}\} \ldots \{\eta^{\pi_{n}},k_{\pi_{n}}\}\right) \right].
\end{split}
\end{equation}
Consider the large $z$ limit of ${\mathcal C}_{\alpha_3}^{\pm}$. The two solutions for the intermediate momenta --- $p^{\pm}$ --- will lead to the same large $z$ scaling and so we will not differentiate between them below. We write
\begin{equation}
\left(p^{\pm}\right)^{\alpha \dot{\alpha}} = \lambda_{31}^{\alpha} \bar{\lambda}_{31}^{\dot{\alpha}}, \quad \left(p^{\pm} + q_1 \right)^{\alpha \dot{\alpha}} = \lambda_{12}^{\alpha} \bar{\lambda}_{12}^{\dot{\alpha}}, \quad \left(p^{\pm} + q_2 \right)^{\alpha \dot{\alpha}} = \lambda_{24}^{\alpha} \bar{\lambda}_{24}^{\dot{\alpha}}
\end{equation}
 At large $z$, we can choose to decompose the  internal momenta so that $\l_{31}, \l_{24}, \lb_{12}$ go large\footnote{Recall that for any decomposition $p_{\alpha \dot{\alpha}} = \l_{\alpha} \lb_{\dot{\alpha}}$ of a momentum vector $p$, we also have the decomposition $p_{\alpha \dot{\alpha}}= (\alpha \l_{\alpha}) {\lb_{\dot{\alpha}} \over \alpha}$ for any complex number $\alpha$. For internal momenta we can choose whatever $\alpha$ we want since this scaling always cancels out in the final answer.} i.e.
\begin{equation}
\label{largezdecsym}
\begin{split}
\l_{31} &= \sum_{n=-1}^{\infty} {\l_{31}^{(n)} \over z^n},  \qquad  \lb_{31} = \sum_{n=0}^{\infty} {\lb_{31}^{(n)} \over z^n},\\
\l_{12} &= \sum_{n=0}^{\infty} {\l_{12}^{(n)} \over z^n},  \qquad  \lb_{12} = \sum_{n=-1}^{\infty} {\lb_{12}^{(n)} \over z^n},\\
\l_{24} &= \sum_{n=-1}^{\infty} {\l_{24}^{(n)} \over z^n},  \qquad  \lb_{24} = \sum_{n=0}^{\infty} {\lb_{24}^{(n)} \over z^n}.
\end{split}
\end{equation}
Momentum conservation gives us important constraints
\begin{equation}
\label{momconstrcut}
\begin{split}
\l_{31}^{(-1)} &= \l_{24}^{(-1)} = \l_{12}^{(0)}, \\
\lb_{31}^{(0)} &= \lb_{24}^{(0)} = \lb_{12}^{(-1)}.
\end{split}
\end{equation}

We are now ready to prove our result. First, we make a change of variables in \eqref{threecut}
\begin{equation}
\eta^{3 1} \rightarrow \eta^{3 1} + \eta^{1 2} z,  \qquad  \eta^{2 4} \rightarrow \eta^{2 4} + \eta^{1 2} z,
\end{equation}
with Jacobian equal to $1$.  The first two tree amplitudes in \eqref{threecut} vanish as ${\rm O}\left({1 \over z}\right)$ by the analysis above. We now make a supersymmetry transformation, on the third tree amplitude,  with parameter
\begin{equation}
\zeta = { (\eta^{2 4} + \eta^{1 2} z)\l_{31} -(\eta^{3 1} + \eta^{1 2} z) \l_{24} \over \<\l_{24},\l_{31}\>}.
\end{equation}
Note that this has the large $z$ expansion $\zeta \xrightarrow[z \rightarrow \infty]{} {\rm O}(1) + {\rm O}\left({1 \over z}\right)$ by \eqref{largezdecsym},\eqref{momconstrcut}. The third tree amplitude now becomes 
\begin{equation}
A^{\mathrm t}(\{0,\l_{24},\lb_{24}\},\{0, \l_{31}, -\lb_{31} \}, \ldots)  = \left(\epsilon_{31}^{-}\right)_{\mu} A^{\mu \nu}  \left(\epsilon_{24}^{-}\right)_{\nu},
\end{equation}
where the two $\epsilon$'s are negative helicity polarization vectors that we can choose to be
\begin{equation}
\label{minpoltrcut}
\left(\epsilon_{31}^{-}\right)^{\alpha \dot{\alpha}} = {\l_{31}^{\alpha} \bar{\mu}^{\dot{\alpha}} \over [ \lb_{31}, \bar{\mu} ]},  \qquad \left( \epsilon_{12}^{-}\right)^{\alpha \dot{\alpha}} = {\l_{12}^{\alpha} \bar{\mu}^{\dot{\alpha}} \over [ \lb_{12}, \bar{\mu} ]}. 
\end{equation}
From the general principles explained in \cite{ArkaniHamed:2008yf}, 
\begin{equation}
\label{generalformlargez}
A^{\mu \nu} \xrightarrow[z \rightarrow \infty]{}{\left({\rm O}(z) + \ldots\right) \eta^{\mu \nu} + \left({\rm O}(1) + \ldots\right) B^{\mu \nu} + \left({\rm O}\left({1 \over z}\right) + \ldots\right)C^{\mu \nu}},
\end{equation}
where $B^{\mu \nu}$ is an anti-symmetric tensor and $C^{\mu \nu}$ is some other tensor.

It is now easy to check using \eqref{momconstrcut},\eqref{minpoltrcut},\eqref{generalformlargez} that the third amplitude grows as ${\rm O}(z)$. However, this means that
\begin{equation}
{\mathcal C}_{\alpha_3}^{\pm} \xrightarrow[z \rightarrow \infty]{} {\rm O}\left({1 \over z}\right).
\end{equation}

As we explained in subsection \ref{findtriangle}, to find the triangle coefficients,
we first need to subtract off the contributions to ${\mathcal C}_{\alpha_3}$ from the box diagrams. The difference between the pure cut and the contribution from the boxes --- that we denote by  ${\mathcal C'}_{\alpha_3}$ ---  must be a polynomial of order $3$ in $z$. In the ${\mathcal N}=4$ theory, the argument above tells us that the cut itself cannot give any contribution to this polynomial. Thus the only non-vanishing contribution must
come, as a remainder, from the higher order terms. The box with an insertion of $p \cdot k$ in the numerator gives a non-vanishing contribution to $C_{\alpha_3}$ even in the limit of large $z$. So, we must have
\begin{equation}
{\mathcal C'}_{\alpha_3} = -\lim_{z \rightarrow \infty} \sum_{\pm}R_{\alpha_3}^{\pm} = B^{(0)}_{\alpha_3},
\end{equation}
with no dependence of $z$ whatsoever! So we see that one triangle coefficient is non-zero even for the ${\mathcal N}=4$ theory but that it is completely determined by the box coefficients. 

A very similar analysis shows us that 
\begin{equation}
{\mathcal C}_2(\omega) \xrightarrow[z \rightarrow \infty]{} {\rm O}\left({1 \over z}\right)
\end{equation} 
However, this time, in the large $z$ limit there is no ${\rm O}(1)$ remainder from the triangles. 
Note that such a remainder would have existed had the coefficient $B_{\alpha_3}^{(m)}$, for $m > 0$ been non-zero. Since this is not the case, from our analysis above, the bubble coefficients must all be zero. 

Thus, we see that in our choice of basis, any one loop amplitude in the noncommutative ${\mathcal N}=4$ theory may be written in terms of the two boxes shown in table \ref{mastertable} and the first triangle on the third line of table \ref{mastertable}. The coefficient of this triangle is completely dictated by the box coefficients. In contrast, in the ordinary ${\mathcal N}=4$ theory --- and in the planar sector of the noncommutative theory --- we can write all amplitudes in terms of the scalar box shown in the first line of table \ref{mastertable}. In a very rough sense, the non-planar sector of the noncommutative theory is about `twice' as complicated as the ordinary theory! We will verify this structure in an example in section \ref{examplesection}.

An interesting implication follows automatically from our result. From our general analysis of non-planar loop diagrams in section \ref{reviewncnp}, we can take a smooth $\theta \rightarrow 0$ limit for the box and triangle integrals that appear in the expansion of the ${\mathcal N}=4$ theory. Hence,  the $\theta \rightarrow 0$ limit of any S-matrix element in the noncommutative ${\mathcal N}=4$ theory gives us the corresponding S-matrix element in the ordinary theory.\footnote{Note that while the $\theta \rightarrow 0$ limit of the S-matrix itself is smooth, by  taking enough derivatives of the S-matrix with respect to $\theta$, one can always arrange for a discontinuity at $\theta = 0$.} This is the statement that there is no UV-IR mixing in the ${\mathcal N}=4$ SYM theory (see \cite{Szabo:2001kg} and references there).

\subsection{Problematizing Simplicity}
\label{problematizesimplicity}
As we have explained above, the S-matrix of the noncommutative ${\mathcal N}=4$ theory is structurally simple, in that only boxes and triangles (whose coefficients are controlled by the boxes) appear in the one-loop S-matrix of this theory. On the surface this would make it seem that ${\mathcal N}=4$ SYM has a simpler S matrix than its non-supersymmetric cousin; this was emphasized in \cite{ArkaniHamed:2008gz}.

However, does this mean that, computationally, the scattering matrix of ${\mathcal N}=4$ is the easiest to obtain? For example, let us say we are calculating gluon scattering at one-loop in a non-supersymmetric theory and in the ${\mathcal N}=4$ theory. Does the use of on-shell techniques make the ${\mathcal N}=4$ calculation easier than the non-supersymmetric calculation? Let us consider the growth of computational complexity with an increasing number of external legs. For simplicity, we will consider planar amplitudes in an ordinary gauge theory but our qualitative conclusions are equally valid for both planar and non-planar amplitudes in the noncommutative theory.

A planar color-ordered amplitude, with $n$ external legs,  receives contributions from ${(n-3)(n-2)(n-1) \over 6}$ boxes, ${(n-2)(n-1) \over 2}$ triangles and $n-1$ bubbles. Thus, for large $n$, most of the work is required to compute the different box coefficients. 

Now, for any box coefficient, we need to compute the product of 3 tree amplitudes and sum over all states that can run in the loop. How does the difficulty of computing a tree amplitude in ${\mathcal N}=4$ SYM compare with the ordinary theory?

If we naively compare the non-supersymmetric recursion relations  \eqref{bcfwrecursion} with their supersymmetric cousins, we would come to the conclusion that tree amplitudes in ${\mathcal N}=4$ SYM are much harder to compute. Each time we cut the amplitude into two parts, we need to sum over $16$ intermediate states in the supersymmetric theory as opposed to $2$ in the ordinary theory. Now, the total computational complexity in computing a $n$ point tree 
amplitude using the BCFW recursion relations, satisfies the 
recursion relation
\begin{equation}
N(n) = 2 g \sum_{j=2}^{n-2} N(j+1),
\end{equation} 
where $g$ is the number of intermediate particles that we need to sum over at each cut i.e. $g=2$ for a pure gauge theory and $g = 16$ for the ${\mathcal N}=4$ SYM theory.
If we set $N(3) = 1$ then for $n > 3$, we have
\begin{equation}
N(n) = 2 g\left(2 g + 1\right)^{(n-4)}, \quad n \geq 4,
\end{equation}
 
 In some cases, this 
formula badly overestimates the complexity of a scattering amplitude in the supersymmetric theory. For example, tree level gluon scattering amplitudes in the supersymmetric theory are exactly the same as in the non-supersymmetric theory.

We do not know how the optimal algorithm for calculating tree amplitudes in ${\mathcal N}=4$ SYM compares with the optimal algorithm for the non-supersymmetric theory. However, it is safe to say that the computation of tree amplitudes in the supersymmetric theory is always {\it at least} as difficult as the non-supersymmetric theory.

Even after this concession, we find that in the supersymmetric theory, to calculate the 4-cut, we need to sum over the entire ${\mathcal N}=4$ multiplet, which contains 16 states, for every cut line. Once again, barring exceptional cases, we always need to do more work for the supersymmetric theory. 

Thus while, for some simple scattering amplitudes involving a small number of external particles, the S-matrix of ${\mathcal N}=4$ SYM might look simpler than that of the non-supersymmetric theory it is, in fact, computationally far more difficult to obtain for a large number of external particles. 

We see then, that while the S-matrix of ${\mathcal N}=4$ SYM is structurally simple, this criterion is mostly aesthetic. From a computational point of view, it still requires more work to compute one-loop scattering in ${\mathcal N}=4$ SYM.

\section{Examples}
\label{examplesection}
We now consider some examples to elucidate the ideas that we have described above. We would like to stress that the methods that we have outlined make the calculation of scattering
amplitudes very simple. For comparison, we invite the reader to compare our calculations in subsection \ref{susyexample} with the usual Feynman diagram calculations for noncommutative ${\mathcal N}=4$ SYM \cite{Liu:2000ad}. 

In our calculations below, we will often draw schematic diagrams of boxes, triangles and bubbles. These diagrams should {\it not} be understood to be Feynman diagrams in any sense. They are merely mnemonics that we use to read off the trace structure and intermediate on-shell conditions when we make cuts. 

In this section, we will drop overall factors of the coupling constant except in subsection \ref{subsecbetafunc}, where we derive the $\beta$ function of Yang-Mills theory.

\subsection{Noncommutative Pure YM: $2 \rightarrow 2$ Scattering}
 
We start with a $2 \rightarrow 2$ scattering amplitude in a $U(N)$ noncommutative gauge theory. This amplitude has a color decomposition, at one-loop, given by \eqref{oneloopcolor}. 
We will work out the coefficient of the ${\rm Tr}(T^1 T^2) {\rm Tr}(T^3 T^4)$ term, where $T^i$ is the color-generator associated with particle $i$.

For convenience, we choose the following initial momenta.
\begin{equation}
\begin{split}
k_1 &= (1,1,0,0),  \quad  k_2 = (-1,1, 0,0), \\
 k_3 &= (-\cosh \phi, -1, 0, -\sinh \phi), \quad k_4 = (\cosh \phi , -1,0, \sinh \phi). 
\end{split}
\end{equation}
We will denote $x \equiv  e^{\phi \over 2}$. We also choose the external helicities to be $h_1 = h_3 = 1, h_2 = h_4 = -1$.
Writing $p_i \sigma^{\mu}_{\alpha \dot{\alpha}} = \left(\lambda_i\right)_{\alpha} \left(\bar{\lambda}_i\right)_{\dot{\alpha}}$, with the Minkowski space condition $\lambda_i^* = \pm \bar{\lambda}_i$, we find the following spinor decomposition of the momenta:
\begin{equation}
\begin{split}
\lambda_1 &= (1,1), \bar{\lambda}_1 =  (1,1); \quad \lambda_2 = (i,-i), \bar{\lambda}_2 =  (i,-i); \\
\lambda_3 &= i (x,1/x), \bar{\lambda}_3 = i  (x,1/x); \quad \lambda_4 = (x,-1/x), \bar{\lambda}_4 =   (x,-1/x).
\end{split}
\end{equation}
Notice that our choice of basis leads to the nice property that $\lambda_i = \bar{\lambda}_i$. This will simplify our calculations a bit below.

We will take $\theta_{23}= -\theta_{32} = {1 \over 2}$ and all other components to be zero.
Equation \eqref{kdef} gives us  $k = \left(0,0,1, 0\right)$.

\subsubsection{Box}
\begin{figure}[!h]
\begin{center}
\scalebox{0.4}{\input{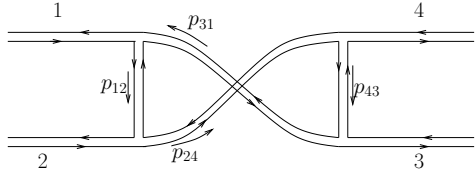}}
\caption{Box: 4-pt YM amplitude}
\label{boxfig}
\end{center}
\end{figure}

For a four-point amplitude, there is a unique partition of the external momenta that can lead to a box and give us the trace structure we want. This is shown schematically in 
 figure \ref{boxfig}. 

The internal momenta $p_{i j}$ that go on shell when we make a four-cut are shown below.\footnote{Note that, in general the momenta $p_{\rm i j}$ are complex. As a result, there is no canonical choice of scaling for their decomposition into spinors. In fact, for each internal momenta, we can choose a convenient scaling because that does not appear in the final answer.}
\[
\begin{array}{l|l|l|l|l}
&p_{31} & p_{12} & p_{24} & p_{43} \\ \hline &&&& \\
p^{+}&{2 i x \over x^2 +  1}  \lambda_1 \bar{\lambda_3} & i {x^2 - 1 \over x^2 + 1} \lambda_1 \bar{\lambda_2} & {2 i x \over 1 + x^2} \lambda_4 \bar{\lambda}_2 &  i {1-x^2 \over 1 + x^2} \lambda_4 \bar{\lambda}_3 \\&&&& \\
p^{-}&{2 i x \over x^2 +  1}  \lambda_3 \bar{\lambda_1} & i {x^2 - 1 \over x^2 + 1} \lambda_2 \bar{\lambda_1} & {2 i x \over 1 + x^2} \lambda_2 \bar{\lambda}_4 &  i {1-x^2 \over 1 + x^2} \lambda_3 \bar{\lambda}_4 \\
\end{array}
\]

Using \eqref{fourcutexplicit}, we find
\begin{equation}
{\mathcal C}_1^{+} = {\mathcal C}_1^{-} = {\left(x^2 - 1\right)^4 \over x^4}.
\end{equation}
This gives us 
\begin{equation}
A_1^{(0)} = {\left(x^2 - 1\right)^4 \over x^4},  \qquad  A_1^{(1)} = 0.
\end{equation}
We always need to consider at least $5$ particles to get a non-zero $A^{(1)}_{\alpha_4}$. We will see an example of this in subsection \ref{susyexample}.

\subsubsection{Triangles}
\begin{figure}[!h]
\begin{center}
\scalebox{0.4}{\input{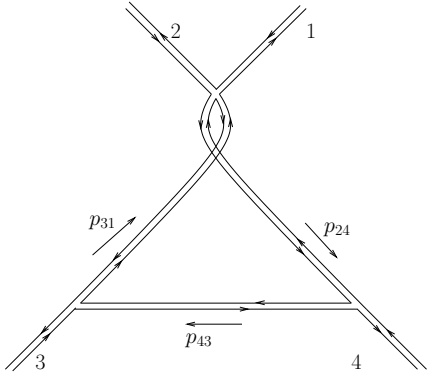}}
\caption{Triangle 1: 4-pt YM amplitude}
\label{triangle12}
\end{center}
\end{figure}

If we want $1,2$ and $3,4$ to appear in distinct traces, only two triangle diagrams are possible. One is where $1,2$ meet at a vertex and the other is where $3,4$ meet at a vertex. Schematically, these two diagrams are shown in figures \ref{triangle12} and \ref{triangle34}.
\begin{figure}[!h]
\begin{center}
\scalebox{0.4}{\input{triangledouble34.pstex_t}}
\caption{Triangle 2: 4-pt YM amplitude}
\label{triangle34}
\end{center}
\end{figure}

\begin{enumerate}
\item {\bf Triangle 1:}
The first triangle is shown in figure \ref{triangle12}. The two internal momenta are obtained by solving \eqref{threeonshell}.
\[
\begin{array}{l|l|l|l}
&p_{31} & p_{24} & p_{43} \\ \hline &&& \\
p^{+}&(z \lambda_4 + \lambda_3) \bar{\lambda}_3 &  \lambda_4 (z \bar{\lambda}_3 - \bar{\lambda}_4) & z \lambda_4 \bar{\lambda}_3  \\ &&& \\
p^{-}&\lambda_3 (-z \bar{\lambda}_4 + \bar{\lambda}_3) & (-z \lambda_3 - \lambda_4) \bar{\lambda}_4 & -z \lambda_3 \bar{\lambda}_4  \\
\hline
\end{array}
\]
We find that
\begin{equation}
\begin{split}
{\mathcal C}^{+}_1  &= -\frac{\left(x^2-1\right)^3   }{x^2 \left(\left(x^2-1\right)   -i \left(x^2+1\right) z\right)}, \\
{\mathcal C}^{-}_1  &= -\frac{\left(i (x^2 + 1) z + (x^2 - 1) \right)^4 + (x^2 + 1)^4 z^4}{x^2 \left(x^2-1\right)  
\left(\left(x^2-1\right) +i \left(x^2+1\right) z\right)}. 
\end{split}
\end{equation}

After subtracting off the remainder from the box terms, as in \eqref{cprimedeftriangle}, we find
\begin{equation}
{\mathcal C}'_1 =  \frac{i \left(x^2+1\right) z \left(-2 \left(x^2-1\right)^2 
   -i \left(x^4-1\right) z   +\left(x^2+1\right)^2
   z^2\right)}{x^2 \left(x^2-1\right) }.
\end{equation}
It is quite remarkable that this term is a polynomial of order $3$ in $z$ but that is exactly what we expect from our general analysis. From here, we can read off
\begin{equation}
\begin{split}
B_1^{(0)} &= 0, \\
B_1^{(1)} &= -\frac{2 i \left(x^2-1\right) \left(x^2+1\right)}{x^2},\\
B_1^{(2)} &= \frac{\left(x^2+1\right)^2}{x^2  }, \\
B_1^{(3)} &= \frac{i \left(x^2+1\right)^3}{x^2 \left(x^2-1\right)  }.
\end{split}
\end{equation}

\item {\bf Triangle 2:}
We now turn to the second triangle diagram shown in \ref{triangle34}. Here the solutions to 
\eqref{threeonshell} are
\[
\begin{array}{l|l|l|l}
&p_{31} & p_{12} & p_{24}  \\ \hline &&& \\
p^{+}&\lambda_1 (-z \bar{\lambda}_2 - \bar{\lambda}_1) & -z \lambda_1 \bar{\lambda}_2 & 
(-z \lambda_1 + \lambda_2) \bar{\lambda}_2 \\&&& \\
p^{-}&(-z {\lambda}_2 - \lambda_1) \bar{\lambda}_1 & -z \lambda_2 \bar{\lambda}_1 & 
\lambda_2 (-z \bar{\lambda}_1 + \bar{\lambda}_2)  \\
\hline
\end{array}
\]
We find that
\begin{equation}
\begin{split}
{\mathcal C}^{+}_2  &= -\frac{\left(-i (x^2 + 1) z + (x^2 - 1) \right)^4 + (x^2 + 1)^4 z^4}{x^2 \left(x^2-1\right)  
\left(\left(x^2-1\right) - i \left(x^2+1\right) z\right)},\\
{\mathcal C}^{-}_2 &= -\frac{\left(x^2-1\right)^3   }{x^2 \left(\left(x^2-1\right) - i \left(x^2+1\right) z\right)}.
\end{split}
\end{equation}
This leads to  
\begin{equation}
{\mathcal C}'_2 = 
\frac{-2 i \left(x^2+1\right) z \left(-2 \left(x^2-1\right)^2 
   +i \left(x^4-1\right) z   +\left(x^2+1\right)^2
   z^2\right)}{x^2 \left(x^2-1\right)  }.
\end{equation}
From here, we can read off
\begin{equation}
\begin{split}
B_2^{(0)} &= 0, \\
B_2^{(1)} &=+\frac{2 i \left(x^2-1\right) \left(x^2+1\right)}{x^2 },\\
B_2^{(2)} &= \frac{\left(x^2+1\right)^2}{x^2  }, \\
B_2^{(3)} &=\frac{-i \left(x^2+1\right)^3}{x^2 \left(x^2-1\right)}.
\end{split}
\end{equation}
\end{enumerate}

\subsubsection{Bubble}
\begin{figure}[!h]
\begin{center}
\scalebox{0.4}{\input{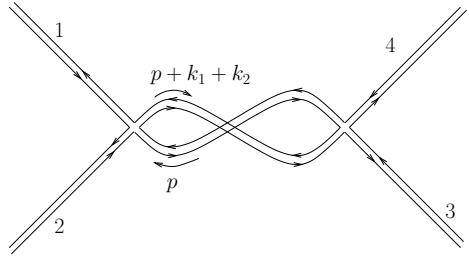}}
\caption{Bubble: 4-pt YM amplitude}
\label{bubblefig}
\end{center}
\end{figure}

There is a single bubble diagram with $1$--$2$ meeting at a vertex and $3$--$4$ meeting at another vertex. This is shown schematically in figure \ref{bubblefig}
We can choose the vectors $w_1, w_2$, described in subsection \ref{subsectionbubblecoefficients}, to be
\begin{equation}
w_1 = (1,0,0,0), w_2 = (0,0,0,i).
\end{equation}
The  solutions to \eqref{bubblecut} become
\begin{equation}
p(\omega) = ( \sqrt{z^2 + 1} \cos \theta, -1, z, i \sqrt{z^2 + 1} \sin \theta).
\end{equation}
Here we have introduced an additional variable $\theta$, with $\omega \equiv e^{i\theta}$ as explained in subsection \ref{subsectionbubblecoefficients}.
The expressions for the intermediate calculations are lengthy, so we just provide the final answer which is
\begin{equation}
\begin{split}
&{1 \over 4 \pi i} \oint_{\omega = \infty} {d \omega \over \omega} \left[{\mathcal C}'(\omega) + {\mathcal C}'\left({1/\omega}\right)\right]\\
&= -\frac{-3 \left(3 x^4-4 x^2+3\right)  +8 i \left(x^2+1\right)^2 z   +\left(7 x^4+16 x^2+7\right) z^2}{8 x^2}.
\end{split}
\end{equation}
From here we can read off
\begin{equation}
\begin{split}
C^{(0)} &= \frac{1}{2} \left(\frac{9 x^2}{4}-3+\frac{9}{4 x^2}\right), \\
C^{(1)} &=0, \\
C^{(2)} &= \frac{-7 x^4-16 x^2-7}{8 x^2  }.
\end{split}
\end{equation}

\subsubsection{Beta Function}
\label{subsecbetafunc}
In the spirit of \cite{ArkaniHamed:2008gz}, we can perform an interesting check on the calculation we have done so far. The idea is as follows. If we combine \eqref{colorordered} and \eqref{oneloopcolor}, we find that the complete
1-loop amplitude for any scattering process has the decomposition
\begin{equation}
\label{treeplusoneloop}
{\mathcal A}^{\mathrm 1 \ell} + {\mathcal A}^{\mathrm t} = \sum_{\pi \in S_n/Z_n} \left(A_{\pi}^{\mathrm t} + N A_{\pi}^{\mathrm p}\right) {\rm Tr}\left(T^{a_{\pi(1)}} \ldots T^{a_{\pi(n)}}\right) + \ldots,
\end{equation}
where the $\ldots$ denote the non-planar terms. In a massless gauge theory, the amplitude above has both UV and IR divergences. We have been working in $4 + 2 \epsilon$ dimensions, but we can trade the dimensional parameter with a running scale $\Lambda$ by performing a $\overline{\mathrm MS}$ renormalization
\begin{equation}
{-1 \over \epsilon} - \gamma + \log(4 \pi) \rightarrow \log \Lambda^2.
\end{equation}.
The scattering amplitude \eqref{treeplusoneloop} now acquires a dependence on $\Lambda$ through the coupling constant and loop integrals and by demanding
\begin{equation}
\label{deriverg}
 \Lambda {d  \left(A_{\pi}^{\mathrm t} + N A_{\pi}^{\mathrm p}\right) \over d \Lambda} = 0,
\end{equation}
we can derive the usual RG equation for the coupling constant. Note that this entire process {\it never} makes any reference to off-shell information! Second, the nonplanar terms in \eqref{treeplusoneloop} do not interfere with the planar terms in \eqref{deriverg}  because  the coefficient of each trace must vanish separately at leading order.

Although we have been considering non-planar amplitudes, we can extract the RG equation from our calculations too. This is 
because the one loop integral coefficients that we have calculated 
are closely related to the loop-coefficients 
of the {\em planar} subamplitude ${A}^{\mathrm p} \left(1, 2, 4, 3\right)$ in an ordinary gauge theory (notice the reversal of the order of $3$ and $4$). As explained in subsection \ref{ordinaryonelooporientation}, we can obtain the loop-coefficients of an ordinary theory by using the noncommutative calculations as an intermediate crutch. Now, the planar subamplitude in the ordinary theory, at one-loop, also receives contributions from other possible partitions of the external momenta.
However, it is easy to check that, with our choice of helicities,  these other partitions never contribute to any ultra-violet divergent terms! 

The ultra-violet divergent terms come from the first and third bubble diagrams and the third triangle diagram in table \ref{mastertable}. The coefficient of ${1 \over \epsilon}$, that we call $\kappa$, is 
\begin{equation}
 32 \pi^2 i \kappa  = 2 C^{(0)} - \left(k_1 + k_2 \right)^2 k^2 {C^{(2)} \over 6} + \left(B^{(2)}_1  + B^{(2)}_2\right) {k^2 \over 2} = {11 \over 3} \left({x^2 \over 2} + {1 \over 2 x^2} - 1\right),
\end{equation}
whereas the tree level amplitude is given by
\begin{equation}
A^{\mathrm t}\left(k_1, k_2, k_4, k_3\right) = {1 \over 2} - {1 \over 4 x^2} - {x^2 \over 4}. 
\end{equation}
If we substitute this into \eqref{deriverg}, and restore factors of the YM coupling constant $g$, we find
\begin{equation}
{d g^2 \over d \ln \Lambda} = {-22 N \over 3} {g^4 \over 16 \pi^2} + \ldots,
\end{equation}
which is the famous RG equation for Yang-Mills theory!

\subsection{Noncommutative ${\mathcal N}=4$ SYM: $2 \rightarrow 3$ Scattering}
\label{susyexample}
Now, we turn to an example of $2 \rightarrow 3$ scattering in noncommutative ${\mathcal N}=4$ SYM.
 For convenience, we take the external momenta to be
\begin{equation}
\begin{split}
k_1 &=  \left(4,4,0,0\right), \quad k_2 = \left(3,0,3,0\right), \quad k_3 = \left(-1,0, -\sin\theta, -\cos\theta\right)\\
k_4 &= \left(-1,0,\sin\theta, \cos\theta,\right), \quad k_5 = \left(-5,-4,-3,0\right) 
\end{split}
\end{equation}
We will consider the scattering of external gluons with helicity $h_1 = h_2 = -1; \, h_3 = h_4 = h_5 = 1$.
We will work with $x = e^{i \theta \over 2}$. We are interested in the coefficient of ${\rm Tr}\left(T^1 T^2\right) {\rm Tr} \left(T^3 T^4 T^5\right)$ in \eqref{oneloopcolor}. We choose $\theta_{42}=-\theta_{24}={-1 \over 4}$ (and all other components zero) so that
$k = \left(0,0,0,1\right)$.

\subsubsection{Boxes}
There are 3 box diagrams that contribute to the part of the amplitude we wish to calculate. We discuss each in turn.
\begin{enumerate}
\item {\bf Box 1}

The first box diagram is shown in figure \ref{boxfig1}. 
\begin{figure}[!h]
\begin{center}
\scalebox{0.4}{\input{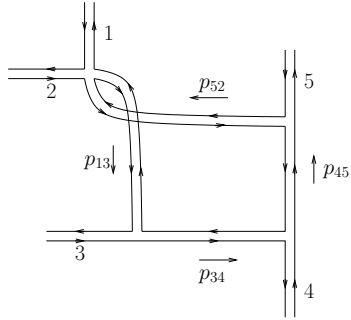}}
\caption{Box 1: 5-pt SYM amplitude}
\label{boxfig1}
\end{center}
\end{figure}
 
There are two possibilities for the momenta to go on shell. These are
\begin{equation}
\begin{split}
p^{+}_{34} &= - {\langle \l_4, \l_5 \rangle \over \langle \l_3, \l_5 \rangle }\l_3 \lb_4,  \\
p^{-}_{34} &= - {\left[ \lb_4, \lb_5 \right] \over \left[ \lb_3, \lb_5 \right]} \l_4 \lb_3.
\end{split}
\end{equation}
Using \eqref{fourcutexplicitsym}, we find
\begin{equation}
\begin{split}
{\mathcal C}^{+}_{1} &= 24 \left({4 \over 5} - {3 i \over 5}\right)\left(1 - 3 i x^2\right), \\
{\mathcal C}^{-}_{1} &= 0.
\end{split}
\end{equation}
Note, that in terms of 
\begin{equation}
\label{mtreesym}
\begin{split}
A^{\mathrm t}\left(2,1,3,4,5\right) &= {\left(\langle \l_2, \l_1 \rangle\right)^3 \over \langle \l_1, \l_3 \rangle \langle \l_3, \l_4 \rangle \langle \l_4, \l_5 \rangle \langle \l_5, \l_2 \rangle} \\
&= {6 i \over 5}{\left(4 - 3 i\right) x^2 \over 3 - i x^2},
\end{split}
\end{equation}
we have
\begin{equation}
\label{boxtreerelation}
{\mathcal C}^{+}_1 = -A^{\mathrm t}(2,1,3,4,5) \left(k_5 + k_4\right)^2 \left(k_3 + k_4 \right)^2.
\end{equation}
This is in complete accordance with  \cite{Cachazo:2008vp}. However, our example differs from the one considered there in two important respects. First, since we are considering the non-planar amplitude, the tree level amplitude that appears in \eqref{mtreesym} is ordered by $\left(2,1,3,4,5\right)$. This switch between the order of $2$ and $1$ also leads to the additional minus sign in \eqref{boxtreerelation}.
Solving \eqref{boxequations}, we find
\begin{equation}
\begin{split}
A_1^{(0)} &= 3 \left({4 \over 5} - {3 i \over 5}\right){\left(x^4 + 9\right)\left(1 - 3 i x^2 \right) \over 1 - x^4},\\
A_1^{(1)} &= -3 \left({3 \over 4} + {4 i \over 5}\right) {\left(3 + i x^2 \right) \left(9 x^4 + 1\right) \over 1 - x^4}.
\end{split}
\end{equation}

\item {\bf Box 2}

As mentioned above, there are two other boxes. 
\begin{figure}[!h]
\begin{center}
\scalebox{0.4}{\input{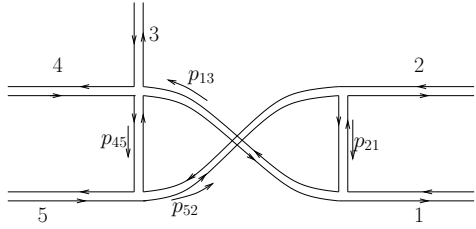}}
\caption{Box 2: 5-pt SYM amplitude}
\label{boxfig2}
\end{center}
\end{figure}
For the box show in figure \ref{boxfig2}, the cut momenta can be
\begin{equation}
\begin{split}
p_{21}^{+} &=  {\langle \l_2, \l_5 \rangle \over \langle \l_1, \l_5 \rangle }\l_1 \lb_2,  \\
p_{21}^{-} &=  {\left[ \lb_2, \lb_5 \right] \over \left[ \lb_1, \lb_5 \right]} \l_2 \lb_1.
\end{split}
\end{equation}
Using the momenta above, we find
\begin{equation}
\begin{split}
{\mathcal C}_2^{+} &= {1728 (3 + 4 i) \over 5} {x^2 \over 3 - i x^2}, \\
{\mathcal C}_2^{-} &= 0.
\end{split}
\end{equation}
which gives us
\begin{equation}
\begin{split}
A_2^{(0)} &= 864 {3 + 4 i \over 5} {x^2 \over 3 - i x^2}, \\
A_2^{(1)} &= 288 {3 i - 4 \over 5} {x^2 \over 3 - i x^2}.
\end{split}
\end{equation}

\item {\bf Box 3}

\nopagebreak
Finally, we come to the third box diagram that is shown in figure \ref{boxfig3}.
\begin{figure}[!h]
\begin{center}
\scalebox{0.4}{\input{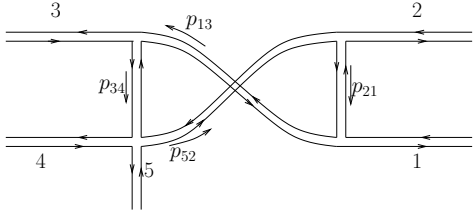}}
\caption{Box 3: 5-pt SYM amplitude}
\label{boxfig3}
\end{center}
\end{figure} 
For this diagram, we have
\begin{equation}
\begin{split}
A_3^{(0)} &= 1152 {3 i - 4 \over 5} {x^4 \over \left(3 - i x^2 \right) \left( 1 + i x^2\right)}, \\
A_3^{(1)} &= 288 { 3 i - 4 \over 5} {x^2 \over 3 - i x^2}.
\end{split}
\end{equation}
\end{enumerate}

\subsubsection{Triangles}
As we have already explained, the triangle coefficients are completely controlled by the box coefficients. However, for this example, we will verify this explicitly. There are three triangle diagrams. 
\begin{enumerate}
\item {\bf Triangle 1}

The first is shown in figure \ref{trsymfig1}.
\begin{figure}[!h]
\begin{center}
\scalebox{0.4}{\input{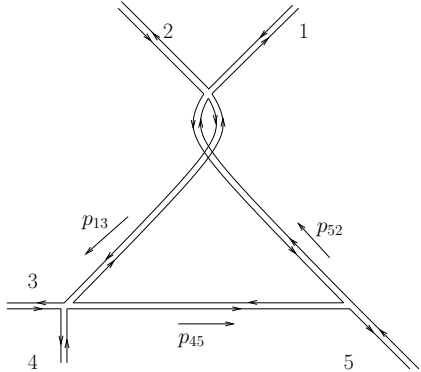}}
\caption{Triangle 1: 5-pt SYM amplitude}
\label{trsymfig1}
\end{center}
\end{figure}
The two momenta that go on shell are
\begin{equation}
\begin{split}
p_{45}^{+} &=\lambda_5 \lb,   \qquad  \lb =  \left({i \over \sqrt{5}} \left(z+1\right), {3 + 4 i \over 5 \sqrt{5}} \left(1 - z \right)\right), \\
p_{45}^{-} &= \lambda \bar{\lambda}_5,  \qquad   \l= \left({i \over \sqrt{5}} \left(z+1\right), {3 - 4 i \over 5 \sqrt{5}}\left(z - 1 \right)\right).
\end{split}
\end{equation}
We find that
\begin{equation}
\label{mtr1}
\begin{split}
{\mathcal C}^{+}_1 &= 240 {x^2 z \left(3 + 4 i \right) \over \left( z - 3 i \right) \left(3 - i x^2 \right) \left[\left(z - 3 i\right) x^2 - 3 i z + 1\right]}, \\
{\mathcal C}^{-}_1 &= 0.
\end{split}
\end{equation}
The first point to notice is that ${\mathcal C}^{+}_1$ goes like ${\rm O}\left({1 \over z}\right)$ for large $z$ which is precisely in accordance with our expectations. However, we would also like to verify that after we have added in the contribution to this cut from the box coefficients we are left with a constant with {\it no} dependence on $z$.
We find that
\begin{equation}
 {\mathcal C}'_1 
=15 {\left(3  + 4 i \right) x^2 \left(x^4 + 1\right) \over \left(3 - i x^2\right) \left(x^4 - 1 \right)}.
\end{equation}
Remarkably, we see that the remainder from the boxes has precisely canceled off the dependence on $z$ in the expression \eqref{mtr1}. From here, we can read off
\begin{equation}
\begin{split}
B_{1}^{(0)} &= 15 {\left(3  + 4 i \right) x^2 \left(x^4 + 1\right) \over \left(3 - i x^2\right) \left(x^4 - 1 \right)}, \\
B_{1}^{(1)} &= 0,  \qquad  B_{1}^{(2)} = 0 \qquad B_1^{(3)} = 0.
\end{split}
\end{equation}

\item {\bf Triangle 2}

An almost identical calculation can be repeated for the triangle shown in figure \ref{trsymfig2} 
\begin{figure}[!h]
\begin{center}
\scalebox{0.4}{\input{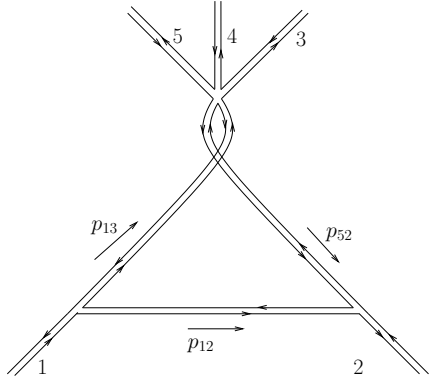}}
\caption{Triangle 2: 5-pt SYM amplitude}
\label{trsymfig2}
\end{center}
\end{figure}. 
We find that
\begin{equation}
\begin{split}
B_{2}^{(0)} &= {-72 \left(3 + 4 i\right) \over 5} {x^2 \left(1 - i x^2 \right) \over \left(3 - i x^2 \right) \left(1 + i x^2 \right)},\\
B_{2}^{(1)} &= 0,  \qquad  B_{2}^{(2)} = 0 \qquad B_{2}^{(3)} = 0.
\end{split}
\end{equation}

\item {\bf Triangle 3}

There is a third triangle shown in figure \ref{trsymfig3}. 
\begin{figure}[!h]
\begin{center}
\scalebox{0.4}{\input{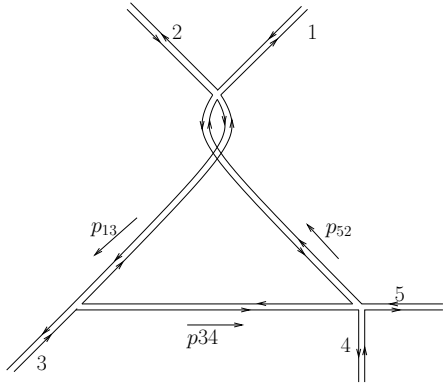}}
\caption{Triangle 3: 5-pt SYM amplitude}
\label{trsymfig3}
\end{center}
\end{figure}
For this final triangle, we find that
\begin{equation}
\begin{split}
B_{3}^{(0)} &= {3\left(4 - 3 i \right) \over 20} {\left(1 - i x^2\right)\left(3 x^4 - 6 i x^2 - 11 \right)\left(11 x^4 - 6 i x^2 - 3 \right) \over \left(1 - x^2\right)\left(1 + x^2 \right) \left(1 + i x^2 \right) \left(3 - i x^2 \right)}, \\
B_{3}^{(1)} &=0,  \qquad  B_{3}^{(2)} = 0 \qquad B_3^{(3)} = 0.
\end{split}
\end{equation}
\end{enumerate}
The coefficients $C^{(m)}$ are all zero as we explained in our general analysis in section \ref{n=4section}.

\section{Results}
Let us briefly recapitulate our results.
\begin{enumerate}
\item
First, we showed that noncommutative tree level amplitudes could be calculated using the BCFW recursion relations. This relies on the remarkable fact that color-ordered amplitudes in noncommutative theories are related to their ordinary counterparts by a simple and calculable phase. Thus, even though the addition of generic higher order terms to the Yang-Mills action makes recursion relations intractable, noncommutative theories --- which contain an infinite number of higher derivative terms --- are amenable to recursion relations.
\item
Second, we showed that one-loop non-planar amplitudes in noncommutative theories could also be calculated via on-shell techniques. We showed that any amplitude could be written as a linear combination of the integrals in table \ref{mastertable}, with coefficients that are rational functions of the external momenta multiplied by a phase that is bilinear in the external momenta. We showed, in section \ref{oneloopsection}, that these coefficients could be efficiently extracted by relating them to products of tree amplitudes. 
\item
We discussed the one-loop S-matrix of the noncommutative ${\mathcal N}=4$ SYM theory and found that it was structurally very simple just like the S-matrix of ordinary ${\mathcal N}=4$ SYM theory. We compared the computational complexity of the supersymmetric theory with the non-supersymmetric theory in subsection \ref{problematizesimplicity}. From a computational point of view, the supersymmetric theory is more expensive.  At one-loop we found that any noncommutative non-planar amplitude in ${\mathcal N}=4$ SYM could be written in terms of the two kinds of boxes in the first two lines of table \ref{mastertable} and the scalar triangle on the third line. However, the coefficient of the triangle was completely determined by the box coefficients. 
\item
Our method of extracting one-loop integral coefficients provides an efficient technique of calculating one-loop integral coefficients in {\it ordinary gauge theories}. This is similar to the technique suggested in \cite{Ossola:2006us}, although our starting point is not an expression for the one-loop integrand but just the on-shell three point amplitude.
\item
In section \ref{examplesection}, we worked out an example of $2 \rightarrow 2$ scattering in noncommutative $U(N)$ YM theory. As part of this calculation we also obtained the $\beta$ function of the ordinary $U(N)$ YM theory. We also worked out a $2 \rightarrow 3$ scattering process in the noncommutative ${\mathcal N}=4$ SYM theory that confirmed our expectations about the structure of the S-matrix for this theory. These calculations are much simpler than the corresponding Feynman diagram calculations.
\end{enumerate}
A natural extension of these ideas would be to understand how well on-shell techniques work at two loops and higher; of course, it would make sense to understand this for ordinary theories first! It would also be very interesting to explore whether such techniques can be used to 
construct new nonlocal perturbative theories.

\section*{Acknowledgments} I would like to thank Nima Arkani-Hamed, Shamik Banerjee, Avinash Dhar, Rajesh Gopakumar, Bobby Ezhuthachan, Shailesh Lal, R. Loganayagam, Shiraz Minwalla, Akitsugu Miwa and Ashoke Sen for helpful discussions.

\appendix
\section*{Appendices}
\section{Reducing Integrals}
\label{reductionnoncommut}
In this appendix, we will show that all non-planar loop integrals that appear in noncommutative theories can be reduced to a set of `master integrals' plus a remainder that contains no branch cut singularities. Our analysis here is similar in spirit to \cite{Ossola:2006us}.

The procedure we discuss below is very similar to the process of reducing one-loop integrals in ordinary theories to boxes, triangles, bubbles and a rational remainder. However, our result for non-planar amplitudes in noncommutative theories differs from the ordinary result in important ways.
\begin{enumerate}
\item
In ordinary theories, all one-loop integrals can be reduced to {\it scalar} integrals with rational coefficients. As we will see, in the noncommutative case, for nonplanar amplitudes, we will also need to include a limited set of tensor integrals. Moreover, the coefficients of this basis are not strictly rational since they include phase factors that are bilinear in the external momenta. These coefficients are, however, free of branch cut singularities. 
\item
In ordinary gauge theories, one-loop integrals are both UV and IR divergent and need to be dimensionally regulated. Rational remainders are obtained when the reduction process is performed carefully within dimensional regularization \cite{Bern:1992em}. 

Non-planar one-loop amplitudes, in a noncommutative gauge theory, have only IR divergences. We will show that these divergences do not cause any subtleties in the reduction process. The 4 dimensional answer is not modified in dimensional regularization. 
\item
Nevertheless, even in the non-planar case, we obtain a remainder that is free of any branch cut singularities. This remainder comes from tadpole graphs that can be ignored in the ordinary theory but not in the noncommutative theory.
\end{enumerate}

Our analysis below is divided into three parts. First, we will show that, in 4 dimensions,  any {\it integrand} that appears in a non-planar one-loop integral may be written as a sum of terms with at most four propagator factors. This is merely an elaborate process of partial fractions. Second, we will show all non-planar integrals can be further reduced to the set of master integrals indicated in table \ref{mastertable}. Finally, we will demonstrate that dimensional regularization does not affect our answer.

\subsection{Four Dimensional Reduction}
In a noncommutative theory, we expect the generic one loop non-planar amplitude
to be a sum over integrals of the form
\begin{equation}
\label{oneloopfraction}
A^{\mathrm np} = \sum \gamma_{\alpha} \int T_{\{r,m\}}^{\alpha} e^{i k \cdot p} \, {d^{4+2\epsilon} p \over (2 \pi)^{4 + 2 \epsilon}},
\end{equation}
where
\begin{equation}
\label{tnmdef}
T_{\{r,m\}}^{\alpha} = {\prod_{j=1}^m 2 (p  \cdot a_j) \over \prod_{i=0}^{r-1} (p + q_i)^2}.
\end{equation}
Note that we have explicitly displayed the dependence 
of $T_{\{r,m\}}^{\alpha}$ on the number of propagator factors $r$ and the number of insertions of $p$ in the numerator, $m$. $T_{\{r,m\}}^{\alpha}$ also depends on the vectors $a_i$ and $q_i$ but we have packaged that dependence in $\alpha$. The $\gamma_{\alpha}$ are some coefficients independent of $p$. We can always choose a gauge so that $m \leq r$ in \eqref{tnmdef} and for simplicity we will assume this below. 

In this subsection, we will assume that the internal momentum $p$ is kept {\it in four dimensions}; this is acceptable since our statements below mostly regard partial fractions. Since we will not deal with integrals in this subsection, we have also dropped the pole prescription in the definition of $T_{\{r,m\}}^{\alpha}$.  We analyze the dimensionally regulated case in subsection \ref{subsecdimreg}.

\subsubsection{Hexagons and Higher Terms}
\label{hexagonshigher}
We start by analyzing terms where the number of propagators in \eqref{tnmdef} is greater than 5. We assume $q_0 = 0$; the generalization to general $q_0$ is obvious. If $q_1, q_2, q_3, q_4$ are linearly independent (this is true for generic momenta), we can write,
\begin{equation}
\label{pdecompose}
p^{\mu} = B_{a b} (p \cdot q_a) q_b^{\mu}.  \qquad  a,b = 1 \ldots 4.
\end{equation}
If $m \geq 1$, the numerator contains the term $p \cdot a_1$ and
\begin{equation}
\label{completesquare}
\begin{split}
2 (p \cdot a_1) &= \sum_{a,b} 2 B_{a b} (p \cdot q_a) (q_b \cdot a_1) \\
&= \sum_{a,b} B_{a b} \left( (p + q_a)^2 - q_a^2 - p^2 \right) (q_b \cdot a_1).
\end{split}
\end{equation}
Inserting this identity into the expression \eqref{tnmdef}, we find
\begin{equation}
T_{\{r,m\}}^{\alpha} = \sum_{\beta}\kappa^{\{r,m\}}_{\alpha, \beta}T^{\beta}_{\{r-1,m-1\}},  \qquad  (m > 1),
\end{equation}
where the $\kappa$ are some coefficients.

If $m = 0$, we choose a non-lightlike $q$, say $q_5$ and write
\begin{equation}
1 = {1 \over q_5^2}\left((p + q_5)^2 - p^2 - 2(p \cdot q_5)\right).
\end{equation}
Inserting this identity into \eqref{tnmdef} and repeating the procedure above, we find
\begin{equation}
T_{\{r,0\}}^{\alpha} = \sum_{\beta} \kappa^{\{r,0\}}_{\alpha,\beta} T_{\{r-1,0\}}^{\beta},
\end{equation}
where the $\kappa$ are some coefficients. 

We can iterate this process till we reach terms with $5$ or fewer propagator factors.

\subsubsection{Pentagons}
\label{pentagonreduction}
For pentagons, which have 5 propagators, if $m > 0$, we can once again use \eqref{pdecompose} and \eqref{completesquare} to reduce the order of the denominator.  If $m = 0$, we write
\begin{equation}
 p^2 = 4 (p \cdot q_a) A_{a b} (p \cdot q_b),  \qquad a,b = 1 \ldots 4
\end{equation} 
where the matrix $A_{a b}$ depends only on the $q$ and not on $p$ which implies that
\begin{equation}
p^2 = \left( (p + q_a)^2 - p^2 - q_a^2 \right) A_{a b} \left( (p + q_b)^2 - p^2 - q_b^2 \right),
\end{equation}
or
\begin{equation}
q_a^2 A_{a b} q_b^2 = -\left((p + q_a)^2 - p^2\right)A_{a b} \left((p + q_b)^2 - p^2\right) + p^2.
\end{equation}
Inserting this in \eqref{tnmdef}, we find an identity of the form
\begin{equation}
T_{\{5,0\}}^{\alpha} = \sum_{\beta} \left(\kappa^{\{5,0\}}_{\alpha, \beta} T_{\{4,0\}}^{\beta} + \kappa'^{\{5,0\}}_{\alpha, \beta} T_{\{4,2\}}^{\beta}\right).
\end{equation}
where the $\kappa, \kappa'$ are some coefficients.

\subsubsection{Boxes and Lower Terms}
\label{boxeslower}
In this subsection, we turn to terms where the number of propagators is $4$ or lower. We consider boxes, triangles and bubbles in turn. 

Boxes have 4 propagators. In place of $q_4$, we now 
use $k$. Of course, we cannot use a $p \cdot k$ term in the numerator to reduce the order of the denominator by $1$. However, using the identity,
\begin{equation}
\label{reducesq}
(p \cdot k)^2  = \# p^2 + (p \cdot k) \#_j  (p \cdot q_j) + \#_{i j} (p \cdot q_i) (p \cdot q_j), \quad i,j = 1 \ldots 3
\end{equation}
where $\#$ are some coefficients, we can ensure that terms quadratic and higher in $p \cdot k$ never appear in the numerator.

We now turn to triangles and bubbles. Our objective is to reorganize these terms into a set of `physical' terms and a set of `spurious' terms that vanish when we consider the integral \eqref{oneloopfraction}. 

To reduce triangles, which comprise terms of the form $T_{\{3,m\}}^{\alpha}$, we introduce an auxiliary vector $v^{\alpha}$, such that $v^{\alpha} \cdot q_i = v^{\alpha} \cdot k = 0$. where $i = 1,2$.  Any 4-vector may be decomposed in terms of $k, q_i, v^{\alpha}$. We still need an identity like \eqref{reducesq}, except we use it to reduce powers of $(p \cdot v^{\alpha})$ to at most linear order. Note the $\alpha$ superscript on $v^{\alpha}$ which serves to remind us that $v^{\alpha}$ depends on $q_i$.

The procedure for reducing bubbles is similar, except we need to introduce two vectors $w_1^{\alpha}$ and $w_2^{\alpha}$ with the property that  $w^{\alpha}_i \cdot q_1 = w^{\alpha}_i \cdot k = 0;\: w_{i}^{\alpha} \cdot w_j^{\alpha} = \delta_{i j}$ (where $i,j$ range over $1,2$). For bubbles, in any case, we can only have up to two insertions of $p$ in the numerator. Furthermore, by symmetry we must have (for some coefficients $\#_i$)
\begin{equation}
p^2 = \#_1 (p \cdot k)^2 + \#_2 (p \cdot q_1)^2 + \#_3 (p \cdot k) (p \cdot q_1) + \#_4 ((p \cdot w_1^{\alpha})^2 + (p \cdot w_2^{\alpha})^2).
\end{equation}
We can use this identity to systematically eliminate the combination $(p \cdot w_1^{\alpha})^2 + (p \cdot w_2^{\alpha})^2$ leaving behind only  $(p \cdot w_1^{\alpha})^2 - (p \cdot w_2^{\alpha})^2$.

To summarize, we have shown that for any fraction of the form $T_{\{r,m\}}$
\begin{equation}
\label{reduce}
\begin{split}
&T_{\{r, m\}}^{\alpha} = \sum_{\beta} \left(c^{\beta}_{4,0} + c^{\beta}_{4,1} (p \cdot k)\right) T_{\{4,0\}}^{\beta} \\
&+  \left[c^{\beta}_{3,0}\left(1 + d_1^{\beta}(p \cdot v^{\beta})\right) + c^{\beta}_{3,1} (p \cdot k)\left(1 + d_2^{\beta}(p \cdot v^{\beta})\right) \right. \\
&\left.+ c^{\beta}_{3,2} (p \cdot k)^2 \left(1 + d_3^{\beta}(p \cdot v^{\beta})\right)+c^{\beta}_{3,3} (p \cdot k)^3 \left(1 + d_4^{\beta}(p \cdot v^{\beta})\right)\right] T_{\{3,0\}}^{\beta} \\
&+ T_{\{2,0\}}^{\beta} \left[
c^{\beta}_{2,0}\left(1 + e_0^{\beta}(p \cdot w^{\beta}_1) + f_0^{\beta}(p \cdot w^{\beta}_2) + g_0^{\beta}\left((p \cdot w^{\beta}_1)^2 - (p \cdot w^{\beta}_2)^2\right) + h_0^{\beta} (p \cdot w^{\beta}_1) (p \cdot w^{\beta}_2)\right) \right. \\
&+ \left. c^{\beta}_{2,1} (p \cdot k) \left(1 + e_1^{\beta}(p \cdot w^{\beta}_1) + f_1^{\beta}(p \cdot w^{\beta}_2) + g_1^{\beta}\left((p \cdot w^{\beta}_1)^2 - (p \cdot w^{\beta}_2)^2\right) +  h_1^{\beta} (p \cdot w^{\beta}_1) (p \cdot w^{\beta}_2)\right) \right. \\
&+\left.  c^{\beta}_{2,2} (p \cdot k)^2 \left(1 + e_2^{\beta}(p \cdot w^{\beta}_1) + f_2^{\beta}(p \cdot w^{\beta}_2) + g_2^{\beta}\left((p \cdot w^{\beta}_1)^2 - (p \cdot w^{\beta}_2)^2\right)+ h_2^{\beta} (p \cdot w^{\beta}_1) (p \cdot w^{\beta}_2)\right)\right] \\
&+ i_1^{\alpha} T_{\{1,0\}} + i_2^{\alpha} T_{\{1,1\}} + i_3^{\alpha} T_{\{0,0\}},
\end{split}
\end{equation}
where the $c,d,e,f,g,h,i$ are some coefficients. So far this is {\it only} a statement about partial fractions in $4$ dimensions. 

In quantum field theory we are interested in loop-integrals. It is easy to see, that when we consider the integral in \eqref{oneloopfraction},
all the terms in \eqref{reduce} that are multiplied by $d, e, f, g,h$ vanish by symmetry. 

\subsubsection{Tadpoles}
Finally, consider the tadpole terms on the last line. Apart from a delta function in $k$ coming from the term $T_{\{0,0\}}$ that we can neglect, we have an integral of the form
\begin{equation}
\int T_{\{1,0\}} e^{i k \cdot p} = \int {e^{i k \cdot p} \over p^2 + i \epsilon} {d^4 p \over (2 \pi)^4} = {1 \over 4 \pi^2 k^2}.
\end{equation}
Hence, the terms on the last line give integrate to give us terms that are free of branch cut singularities. These rational terms were discussed in \cite{Hayakawa:1999zf,Matusis:2000jf}
\subsubsection{Conclusion}
We will show in the next subsection that the conclusions of the previous subsections are unaffected when we work within dimensional regularization. This leaves us with the result that any noncommutative non-planar one-loop amplitude may be written as
\begin{equation}
\label{decompositionnoncommutappendix}
\begin{split}
A^{\mathrm np}_{j;\pi} &=  \sum_{{\alpha_4}} \int {\sum_{m = 0}^{1} A_{\alpha_{4}}^{(m)} (p \cdot k)^m e^{i p \cdot k} \over \prod_{i = 0}^{3}\left[(p + q^{\alpha_{4}}_i)^2 + i \epsilon\right]} \, {d^{4+2\epsilon} p \over (2 \pi)^{4 + 2\epsilon}} \\
&+ \sum_{\alpha_{3}} \int {\sum_{m = 0}^{3} B_{\alpha_{3}}^{(m)}(p \cdot k)^m  e^{i p \cdot k} \over \prod_{i = 0}^{2}\left[(p + q^{\alpha_{3}}_i)^2+i \epsilon \right]} \, {d^{4+2\epsilon} p \over (2 \pi)^{4 + 2\epsilon}} \\
&+  \int {\sum_{m = 0}^{2} C^{(m)}(p \cdot k)^m  e^{i p \cdot k} \over \prod_{i = 0}^{1}\left[(p + q_i)^2 + i \epsilon\right]} \, {d^{4+2\epsilon} p \over (2 \pi)^{4 + 2\epsilon}}\\
&+ {\mathcal R} + {\rm O}\left(\epsilon\right),
\end{split}
\end{equation}
where ${\mathcal R}$ is a remainder with no branch cut singularities, as explained above. The equation above is true as a power series in $\epsilon$ up to and including terms of $O(\epsilon^0)$. We will now justify this statement.

\subsection{Dimensional Regularization}
\label{subsecdimreg}
Noncommutative non-planar one-loop amplitudes suffer from IR-divergences. To regulate the theory, we work within dimensional regularization. It is convenient to break up the loop momentum into a four dimensional part and a $2 \epsilon$ dimensional part \cite{Bern:1995db,Mahlon:1993si} (see also \cite{Badger:2008cm} and references there)
\begin{equation}
p_d = p_4 + \mu_{2 \epsilon},
\end{equation}
where $d=4+2\epsilon$. In the four dimensional helicity scheme \cite{Bern:1991aq}, the external momenta and polarization vectors are always kept in 4 dimensions. Thus, if $q_i, a_j$ are external vectors 
\begin{equation}
\begin{split}
(p_d + q_i)^2 &= p_4^2 + 2 p_4 \cdot q_i + q_i^2 + \mu_{2 \epsilon}^2 \\
p_d \cdot a_j &= p_4 \cdot a_j
\end{split}
\end{equation}
To lighten the notation, we will now drop the $2 \epsilon$ subscript under $\mu$. 

It is evident, that in this scheme, the calculation of subsection \ref{hexagonshigher} go through without change. However, if we redo the calculations of subsection \ref{pentagonreduction} and \ref{boxeslower} carefully we find that (apart from `spurious terms' that integrate to zero by symmetry) we get additional 
terms of the form
\[
{ {\mu}^{2 j} (p_4 \cdot k)^{j_1} \over  \prod_{i=0}^r \left[(p + q_i)^2+ i \epsilon\right]},
\]
where $r \leq 4$.
The integral over all such terms can be obtained from the integral 
\begin{equation}
\label{remainderprototype}
I_R =i \int {\mu^{2 j}  e^{i \alpha  k \cdot p_4}   \over  \prod_{i = 0}^{r} \left[(p_4 + q_i)^2 - \mu^2 + i \epsilon\right]} \, {d^4 p_4 \over (2 \pi)^4} {d^{2 \epsilon} \mu \over (2 \pi)^{2 \epsilon}},
\end{equation}
by taking derivatives with respect to $\alpha$ and then setting $\alpha = 1$. 
After combining denominators using Feynman parameters and Wick rotating, we find the integral is
\begin{equation}
\begin{split}
I_R &= {(-1)^r \over (2 \pi)^{4 + 2 \epsilon}} \int d x_F \, d^4p\, d^{2 \epsilon}\mu \, {\mu^{2 j} e^{i \alpha p_E' \cdot k} \over (p_E^2 + \Delta_0 + \mu^2)^{r+1}} \\
&=  {(-1)^r \over (2 \pi)^{4 + 2 \epsilon} \Gamma(r+1)} \int  d x_F \,d^4p \,d^{2 \epsilon}\,\mu d \beta \left[ \mu^{2 j} {e^{i \alpha p_E' \cdot k - \beta (p_E^2 + \Delta_0 + \mu^2)} \beta^{r}}\right] \\
&=   {(-1)^r  \pi^{\epsilon} \over  (2 \pi)^{4+2\epsilon} \Gamma(r+1) \Gamma(\epsilon)}
\int d x_F \, d^4 p \, d(\mu^2) \, d \beta \left[ \,(\mu^2)^{j + \epsilon - 1} {e^{i \alpha p_E' \cdot k - \beta (p_E^2 + \Delta_0 + \mu^2)} \beta^{r}}\right],
\end{split}
\end{equation}
where we have introduced the notation
\begin{equation}
\begin{split}
\Delta_0 &= -\sum_i q_i^2 x_i + \left(\sum_i q_i x_i\right)^2,\\
d x_F &= \prod d x_i \delta\left(1 - \sum x_i\right), \quad p_E' = p_E - \sum q_i x_i.
\end{split}
\end{equation}

We now do the integral over $\mu, p_E, \beta$ in that order.
\begin{equation}
\label{intwithmu}
\begin{split}
I_R &= {(-1)^r \Gamma(j + \epsilon) \pi^{\epsilon} \over  (2 \pi)^{4 + 2 \epsilon} \Gamma(r+1) \Gamma(\epsilon)}
\int d x_F \,d^4p \,d\beta \left[{e^{i \alpha p_E' \cdot k - \beta (p_E^2 + \Delta_0)} \beta^{r - j - \epsilon}}\right] \\
&= {(-1)^r \Gamma(j + \epsilon) \pi^{2+\epsilon} \over  (2 \pi)^{4 + 2 \epsilon} \Gamma(r+1) \Gamma(\epsilon)}
\int d x_F \,d\beta \left[e^{{\alpha^2 k^2 \over 4 \beta} - \beta \Delta_0}  \beta^{r-2 - j - \epsilon}  e^{-i \alpha \sum x_i q_i \cdot k}\right] \\
&=  {(-1)^r \Gamma(j + \epsilon) \over 2^{r-j} (2 \pi)^{2 + \epsilon} \Gamma(r+1) \Gamma(\epsilon)}
\int d x_F\left[{ (\alpha |k|)^{r_0} \over \Delta_0^{r_0 \over 2}} K_{r_0}(\alpha |k| \sqrt{\Delta_0})  e^{-i \alpha \sum x_i q_i \cdot k}  \right], 
\end{split}
\end{equation}
where $r_0 = r - j - 1 - \epsilon$ and $|k| = \sqrt{-k^2}$ (recall that the dot product is taken with the Lorentzian metric, so $|k|$ is always real). It is easy to check that for $j =0, \alpha = 1$, the answer in \eqref{intwithmu} reduces to \eqref{ncnpbessel}.

If we use the small parameter expansion $K_{r_0}(x) \sim {2^{|r_0|-1}\Gamma(|r_0|) \over x^{|r_0|}}, x << 1$ it is possible to check that the integral over the Feynman parameters has no ${1 \over \epsilon}$ pole for $j \geq 1$. Hence, the prefactor  ${1 \over \Gamma(\epsilon)}$ multiplying the integral ensures that it vanishes in the limit $\epsilon \rightarrow 0$. When we differentiate with respect to $\alpha$, we use
\begin{equation}
{\partial \over \partial \alpha} K_{r_0}(\alpha k \sqrt{\Delta}) = {-k \sqrt{\Delta} \over 2}\left( K_{r_0+1}(\alpha k \sqrt{\Delta}) +  K_{r_0-1}(\alpha k \sqrt{\Delta}) \right).
\end{equation}
 So, even derivatives of \eqref{remainderprototype} with respect to $\alpha$ vanish in the limit $\epsilon \rightarrow 0$, which proves our result.

\bibliographystyle{JHEP}
\bibliography{references}
\end{document}